\documentclass{aa}  

\usepackage{graphicx}
\usepackage{txfonts}
\usepackage{color}

\usepackage{ulem}
\usepackage{tablefootnote}
\usepackage{float} 
\usepackage{longtable}
\usepackage[labelfont=small,bf]{caption}
\usepackage{amsmath}
\usepackage[english]{babel}
\usepackage{comment}
\usepackage{wasysym}
\usepackage{hyperref}
\hypersetup{%
  colorlinks = true,
  linkcolor  = black
}

\hypersetup{colorlinks=true,citecolor=blue}

\usepackage{natbib}
\bibpunct{(}{)}{;}{a}{}{,} 

\begin{document}

   \title{Wide companions to M and L subdwarfs with Gaia and the Virtual Observatory\thanks{Tables A-1, A-2, and A-3 are only available in electronic form at the CDS via anonymous ftp to \url{cdsarc.u-strasbg.fr} (130.79.128.5) or via \url{http://cdsweb.u-strasbg.fr/cgi-bin/qcat?J/A+A/}}}


   \author{J.\ Gonz\'alez-Payo \inst{1,2}, M.\ Cort\'es-Contreras \inst{2,3}, N.\ Lodieu \inst{4,5}, E.\ Solano \inst{2,3}, Z.\ H.\  Zhang \inst{6}, M.-C.\ G\'alvez-Ortiz \inst{7}}

   \institute{Departamento de F\'isica de la Tierra y Astrof\'isica. Universidad Complutense de Madrid (UCM), Plaza de Ciencias 3, E-28040, Madrid, Spain.
         \email{fcojgonz@ucm.es}
         \and
         Departamento de Astrof\'isica, Centro de Astrobiolog\'ia (CSIC-INTA), ESAC Campus, Camino Bajo del Castillo s/n, E-28692, Villanueva de la Ca\~nada, Spain
         \and
         Spanish Virtual Observatory, Spain  
         \and
         Instituto de Astrof\'isica de Canarias (IAC), Calle V\'ia L\'actea s/n, E-38200 La Laguna, Tenerife, Spain. 
         \and
         Departamento de Astrof\'isica, Universidad de La Laguna (ULL), E-38205 La Laguna, Tenerife, Spain.
         \and
         School of Astronomy and Space Science, Key Laboratory of Ministry of Education, Nanjing University, 163 Xianlin Avenue, Nanjing 210023, China
         \and
         Departamento de Econom\'ia Financiera y Contabilidad e Idioma moderno. Facultad de Ciencias Jur\'idicas y Sociales. Universidad Rey Juan Carlos (URJC). Campus de Vic\'alvaro, E-28032, Madrid, Spain
             }

   \date{Received February 3, 2021; accepted April 19, 2021}
 
  \abstract
   {}
   {The aim of the project is to identify wide common proper motion companions to a sample of spectroscopically confirmed M and L metal-poor dwarfs (also known as subdwarfs) to investigate the impact of metallicity on the binary fraction of low-mass metal-poor binaries and to improve the determination of their metallicity from the higher-mass binary.}
   {We made use of Virtual Observatory tools and large-scale public surveys to look  in \textit{Gaia}  for common proper motion companions to a well-defined sample of ultracool subdwarfs with spectral types later than M5 and metallicities below or equal to $-$0.5\,dex. We collected low-resolution optical spectroscopy for our best system, which is a binary composed of one sdM1.5 subdwarf and one sdM5.5 subdwarf located at $\sim$1\,360\,au, and for another two likely systems separated by more than 115\,000\,au.}
   {We confirm one wide companion to an M subdwarf, and infer a multiplicity for M subdwarfs (sdMs) of $1.0_{-1.0}^{+2.0}$\% for projected physical separations of up to 743\,000\,au. We also find four M--L systems, three of which are new detections. No colder companion was identified in any of the 219 M and L subdwarfs of the sample, mainly because of limitations on the detection of faint sources with \textit{Gaia}. We infer a frequency of wide systems for sdM5--9.5 of $0.60_{-0.60}^{+1.17}$\% for projected physical separations larger than 1\,360\,au (up to 142\,400\,au). This study shows a multiplicity rate of $1.0_{-1.0}^{+2.0}$\% in sdMs, and $1.9_{-1.9}^{+3.7}$\% in extreme M subdwarfs (esdMs). We did not find any companion for the ultra M subdwarfs (usdMs) of our sample, establishing an upper limit of 5.3\% on binarity for these objects.}
   {}

   \keywords{subdwarfs --- stars: low mass --- techniques: photometric --- astrometry --- surveys --- virtual observatory tools}

  \authorrunning{J.\ Gonz\'alez-Payo}
  \titlerunning{Wide companions to M and L subdwarfs with Gaia and the Virtual Observatory}

   \maketitle

\section{Introduction}
\label{SpClass_sdM:intro}
Subdwarfs are objects that lie appreciably below the main sequence on the Hertzsprung-Russell diagram and were first discovered by \citet{kuiper39}. They have a luminosity class VI under the Yerkes spectral classification system \citep{morgan43}, and appear less luminous than solar metallicity dwarfs with similar spectral types because of the low abundances in elements heavier than helium. Subdwarfs belong to population II and are stars from the galactic thick disk or halo \citep{gizis99}.

Cool subdwarfs have spectral types G, K, and M, and are typically found to have thick disk or halo kinematics \citep{gizis97a}. They are presumably relics of the early Galaxy, with ages of 10$-$12\,Gyr \citep{jofre11}, and are therefore excellent tracers of Galactic chemical history, because they were formed at the early stage of the Milky Way. 

There are different metallicity classes of M subdwarfs based on a spectral index that measures the ratio of hydrides and oxides present in their atmospheres. The original classification for M subdwarfs (sdMs) and extreme subdwarfs (esdMs) developed by \citet{gizis97a} was revised and extended by \citet{lepine07c}. A new class of subdwarfs, the ultra subdwarfs (usdMs), has been added to the sdMs and esdMs. Currently, dwarfs are classified as having solar metallicity, subdwarfs as having  moderately low metallicity, extreme subdwarfs as having very low metallicity, and ultra subdwarfs as having ultra low metallicity \citep{kirkpatrick05} with metallicity estimates of approximately $[M/H]\,=\,0$, $-0.5$, $-1.0$, and $-2.0$ dex, respectively, with a typical dispersion of 0.5 dex \citep{lodieu17a,zhang17a}.

Star-like objects with spectral types later than M5 and effective temperatures of less than $\sim$2900\,K \citep{kirkpatrick97b} are usually referred to as ultracool dwarfs. This heterogeneous group includes stars of extremely low mass as well as brown dwarfs, and represents about 15\% of the population of astronomical objects near the Sun \citep{gillon16}. The M dwarfs represent about two-thirds of the stars in the Milky Way and constitute around the 40\% of the total stellar mass in the Galaxy \citep{gould96,bochanski10}.

M dwarfs are now of popular interest in the search for extrasolar planets with complementary techniques, leading independent groups to define new methods to estimate their metallicities. As of now, most studies look at slightly metal-poor M dwarfs (typically $>$\,$-$0.5 dex) with an accuracy in their metallicity determination of the order of 0.15 dex, either photometrically \citep{bonfils05a,johnson09a,schlaufman10a,neves12,hejazi15,dittmann16}, or spectroscopically \citep{woolf05,woolf06,bean06a,woolf09,rojas_ayala10,rojas_ayala12,muirhead12,terrien12a,onehag12,neves13,neves14,hejazi15,newton15a,lindgren16a}. The sample of M dwarfs with metallicities of less than $-$0.5 dex is small, and very few M dwarfs in that sample have companions to more massive primaries with well-determined physical parameters to help to determine the metallicity scale of sdMs.

There is a direct correlation between the metallicity of stars and the occurrence of giant gaseous exoplanets \citep{papaloizou06,williams11}. In the core accretion model of planet formation \citep{pollack96,papaloizou06,udry07,boley09,janson11}, there are particular processes that depend on metallicity to form a planet from a dusty circumstellar disk \citep{weidenschilling80,armitage10}. Those processes tend to occur in greater numbers in metal-rich disks than in metal-poor disks \citep{johnson12}. Therefore, it is of prime importance to determine the metallicity of M dwarfs with the best accuracy possible in order to better characterise the properties of the planets in their vicinity.

In this paper, we present a dedicated search for wide companions to known M and L subdwarfs reported in the literature \citep{lodieu12b,lodieu17a,zhang17b,zhang17a,zhang18b,zhang18a}. The objective of this paper is two-fold: (i) to determine the multiplicity rate of our sample, and (ii) to improve estimates of the metallicity of M/L subdwarfs and their distances from their more massive primary. We look for wide common proper motion companions in \textit{Gaia} DR2, and we compare our results with multiplicity studies focusing on metal-poor populations such as those of \cite{chaname04},  \cite{zapatero04a}, \cite{riaz08a}, \cite{jao09a}, \cite{badenes18a}, \cite{moe19}, and \cite{El_Badry19a}.

The paper is structured as follows. In Section~\ref{wide_sdM:sample}, we describe the sample of M and L subdwarfs and their properties. In Section~\ref{wide_sdM:method}, we describe the methodology used to identify wide common proper motion companions on the basis of astrometric and photometric criteria. In Section~\ref{wide_sdM:analysis}, we analyse all potential companion candidates based on their proper motions, distances, and photometry. In Section~\ref{wide_sdM:results_search}, we present the results of our search and the most promising wide systems with additional photometric and spectroscopic characterisation. In Section~\ref{wide_sdM:discuss}, we provide the spectral type of each component of the multiple systems identified in our search, and we discuss the frequency of M and L subdwarf systems with previous observational studies and theoretical predictions. 

\section{Sample selection}
\label{wide_sdM:sample}

To achieve our scientific objectives, we worked with a sample of 219 known ultracool subdwarfs, 185 of which were taken from the SVO late-type subdwarf archive\footnote{\url{http://svo2.cab.inta-csic.es/vocats/ltsa/index.php}} maintained by the Spanish Virtual Observatory\footnote{\url{https://svo.cab.inta-csic.es/main/index.php}}. The sample of subdwarfs contained in the archive is an extension of the list of 100 subdwarfs identified by \citet{lodieu17a} using Virtual Observatory (VO) tools, now containing 193 sources. This sample includes most of the known ultracool subdwarfs confirmed spectroscopically at the time of writing. We reject eight of them because of the lack of spectral types derived from optical spectroscopy. For each object, the archive contains coordinates, identifiers, effective temperatures, proper motions, spectral types, and magnitudes in different passbands, which can be accessed through a very simple search interface that permits queries by coordinates or radius and/or a range of magnitudes, colours, and effective temperatures. All of these data come from the Two Microns All Sky Survey \citep[2MASS;][]{cutri03}, the United Kingdom InfraRed Telescope (UKIRT) Deep Sky Survey \citep[UKIDSS;][]{lawrence07}, the Visible and Infrared Survey Telescope for Astronomy (VISTA) Hemisphere Survey \citep{mcmahon12}, the Sloan Digital Sky Survey \citep[SDSS;][]{york00} and Wide Field Infrared Survey Explorer \citep[WISE;][]{wright10} public catalogues. Thirty-four additional objects were taken from more recent works \citep{zhang17b,zhang17a,zhang18b,zhang18a,zhang19g}, bringing the total sample used here to 219 ultracool subdwarfs. This sample contains all known subdwarfs with spectral types between M5 and L7 confirmed spectroscopically.

The 219 sources considered in this work cover spectral types from M5 to L8, and belong to different metallicity classes: subdwarfs, extreme subdwarfs, and ultra subdwarfs. The numeric identifiers (Id) of the 219 sources, common names, coordinates, source identification from $Gaia$ DR2 \citep{Gaia_Brown2018} when existing, spectral types, and references are listed in Table~A-1, and can  also be retrieved from \textit{VizieR} \citep{ochsenbein00}. Coordinates were obtained from the catalogue that gives the name to the source: SDSS \citep{adelman_mccarthy09,adelman_mccarthy12,alam15a}, ULAS \citep[UKIDSS Large Area Survey;][]{lawrence07}, 2MASS \citep{cutri03}, LSR \citep[L\'epine, Shara, Rich;][]{lepine02}, LHS \citep[Luyten Half Second catalogue][]{luyten79}, SSSPM \citep[SuperCOSMOS Sky Survey Proper Motion;][]{hambly01a}, and APMPM \citep[Automatic Plate Measuring Proper Motion;][]{kibblewhite71}.

\section{Methodology}
\label{wide_sdM:method}
In this section, we describe our methodology using proper motion, distance, and binding energy criteria to identify wide companions to our sample of M and L subdwarfs. We based all of our data on catalogue \textit{Gaia} DR2, and we did not use the latest  release of $Gaia$ because we started this work well ahead of the $Gaia$ EDR3\@.

\subsection{Proper motions}
\label{wide_sdM:method_PM}
To search for wide companions, we need the most accurate proper motions possible. To collect them, we looked for proper motions in the $Gaia$ DR2 catalogue. Whenever proper motions were not available, we computed them through a linear regression of the positions and epochs provided by different astrometric catalogues: 2MASS, SDSS DR7 \citep{adelman_mccarthy09}, SDSS DR9 \citep{adelman_mccarthy12}, SDSS DR12 \citep{alam15a}, UKIDSS DR9 \citep{lawrence07}, WISE \citep{wright10}, and Pan-STARRS1 \citep{chambers16a}.
We used the \textit{Aladin} Sky Atlas \citep{bonnarel00} VO tool, \textit{Simbad} \citep{wenger00}, and \textit{VizieR} VO services to avoid mismatches and identify the correct detections of the sources in our sample from each catalogue.
We checked whether or not the method agrees with \textit{Gaia} DR2 values when these latter were available, and we found variations of less than 4\% in right ascension and declination components and less than 10\% in the total proper motion.

Using the linear regression method, for most of the sources we found between 4 and 15 positions with different epochs, even in the same catalogue. The mean coverage is around 10 years, giving acceptable error bars. For the 219 M and L subdwarfs in this work, we have 149 with proper motions from $Gaia$ DR2 and 70 computed by linear regression using public catalogues. The median motion of the objects in our sample is around 255 mas/a. The mean uncertainty on the proper motions is about 1 mas/a and 12 mas/a for those in $Gaia$ DR2 and those computed by linear regression, respectively. We calculated the proper motion with just two values for two sources, yielding large errors in those specific cases (Id 29 and Id 134). We did not find any companions to these two sources even after applying such error margins. We list the proper motions and their references in the second, third, and fourth columns of Table~A-2. The data contained in this table can be retrieved from \textit{VizieR}.

\subsection{Distances}
\label{wide_sdM:method_dist}
We also need distances that are as accurate as possible. Therefore, we first considered the parallaxes from $Gaia$ DR2. If not available or if the parallax error was higher than 20\% of the parallax, we estimated spectrophotometric distances as in \citet{lodieu17a}.
Although \citet{bailerjones18a} set a limiting value of the relative error of the parallax at 10\%, we allow a larger margin (20\%) to avoid rejecting possible candidates (e.g. unresolved binaries). Because of the weakness of the sources, the uncertainties on \textit{Gaia} DR2 parallaxes are similar to the uncertainties on the calculated distances. We found 126 M and L subdwarfs with $Gaia$ DR2 parallaxes and estimated spectro-photometric distances for another 93 objects in our sample. In some cases, the $Gaia$ distance with errors on the parallax higher than 20\% turns out to be more accurate than that estimated from the spectral type--magnitude relation. We therefore opted for the $Gaia$ distances in those specific cases.

To calculate the spectrophotometric distances, we used the \textit{J} band photometry in Table~5 in \citet{lodieu17a} when available, and the photometry in the \textit{i} and \textit{J} bands in Table~2 in \citet{zhang13} otherwise. The error on the spectrophotometric distances takes into account the 0.5 uncertainty on the spectral type and the associated error on the magnitudes, and other parameters used in the relation given in \citet{zhang13}.

\begin{figure}
  \centering
  \includegraphics[width=1\linewidth, angle=0]{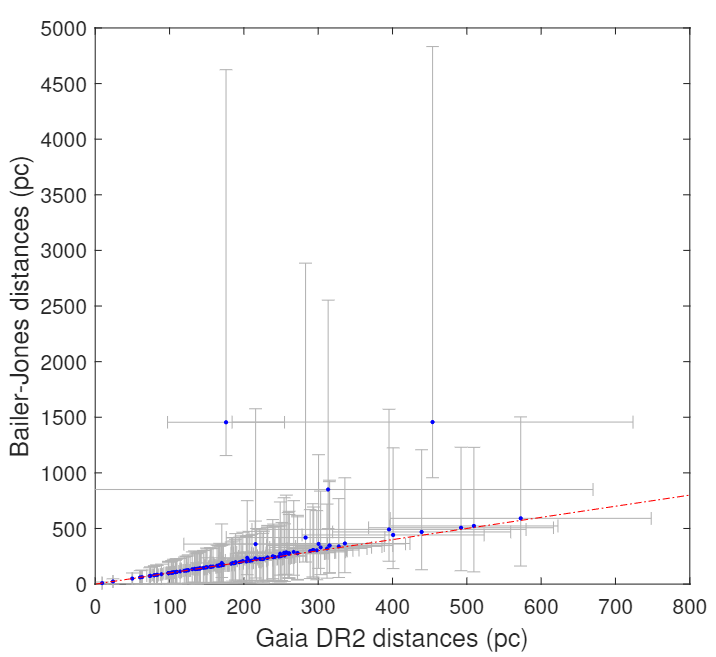}
     \caption{Comparison between $Gaia$ DR2 distances and the ones in \citet{bailerjones18a}. The red dashed line represents the 1:1 relation. 
}
   \label{fig_wide_sdM:Bailer_Jones}
\end{figure}

The simple approach of inverting the parallax to estimate the distance can sometimes lead to potentially strong biases, especially (but not only) when the relative uncertainties are large and objects lie at large distances, as is often the case for members of the halo. A proper statistical treatment of the data, its uncertainties, and correlations may be required as advised by \citet{luri18a} and \citet{bailerjones18a}. For this reason, we compared the adopted $Gaia$ DR2 distances in this work with the values of \citet{bailerjones18a}, who recently proposed an alternative methodology to get reliable distances taking into account the non-linearity of the transformation and the asymmetry of the resulting probability distribution. In our case, although the $Gaia$ DR2 distances are not reliable when the error on the parallax is larger than or equal to 20\%, the comparison with the distances from \citet{bailerjones18a} shows that these distances have larger errors than the $Gaia$ DR2 ones (see Fig.~\ref{fig_wide_sdM:Bailer_Jones}). We observe that 5 sources out of 123 lie noticeably away from the 1:1 relation, most of them showing unreliable distances and very large errors. We also note consistency between our spectrophotometric distances and the ones in $Gaia$ DR2, when available. Therefore, the use of the distances from \citet{bailerjones18a} does not provide any significant advantage with respect to the use of $Gaia$ DR2 or spectrophotometric distances. Finally, we were able to compile and compute distances for all the sources in our sample, which are presented in the fifth and sixth columns in Table~A-2. The mean distance of the sample is 187\,pc with a mean error of 18.1\,pc in $Gaia$ distances and 33.7\,pc in the spectrophotometric distances.

\subsection{Radius of the search}
\label{wide_sdM:method_radii}
To perform the search for companions, we define a search radius for M and L subdwarfs equal to the maximum separation the system may have if gravitationally bound. Beyond that separation, the gravitational binding energy is too low to keep the system tight and pairs are no longer physically bound. This search radius is determined as a function of the binding energy ($W$) and the masses involved in the system:

\begin{equation}
W=G\frac{m_1 m_2}{r}
\label{eqn:binding}
,\end{equation}

\noindent where $G$ is the gravitational constant and has a value of $6.674\cdot 10^{-11}$ \:Nm$^2$kg$^{-2}$ \citep{carroll07}, $m_1$ and $m_2$ are the masses in kilograms of the two components of the system, and $r$ is the projected physical separation between them in metres. The most commonly accepted value of the minimum energy required for two celestial bodies to be bound is $10^{33} J$ \citep{caballero09,dhital10b}. Accounting also for the maximum mass that a physical companion could have, we can obtain the maximum separation between components.

We set A0 to be the upper limit in the spectral type (i.e. in the mass) of the companion. Although lifetimes of A0 stars are typically shorter than lifetimes of subdwarfs, we select them in order to account for the extra mass involved in close binaries, that is,\ in the case of triple or multiple systems, that could yield larger projected physical separations. This maximum separation is calculated for each source in the sample using the average mass of an A0 star and the mass of the subdwarf. We estimated a mass of 2.36$\pm$0.035\,$M_\odot$ for an A0 star based on the data from \citet{popper80}, \citet{harmanec88}, and \citet{graydf05}. We adopted a mass of $2.40\,M_\odot$ accounting for the estimated error, which is in agreement with \citet{adelman04}. The masses of the low-mass stars vary depending on their spectral types and metallicity \citep{burrows89,kroupa97}. Because of the lack of dynamical masses for metal-poor ultracool dwarfs, we adopted the masses of main sequence solar-type M dwarfs from \citet{reid05b} as a first approximation. For stars with spectral types later than M9, we used the mass of 0.075 M$_{\odot}$ as an upper limit, corresponding to the stellar--substellar boundary at solar-metallicity \citep{chabrier97}.

\subsection{Search criteria}
\label{wide_sdM:search_criteria}
Once we inferred the proper motion, distance, and search radius for each source in our sample, we looked for candidate companions in $Gaia$ DR2 through TOPCAT \citep{taylor05} and a code in ADQL\footnote{Astronomical Data Query Language.} specifically written for our purposes \citep{yasuda04}. We imposed the following conditions, where \textit{Obj} refers to the subdwarf in our sample and \textit{Comp} to the companion candidate:

\begin{itemize}
\item [$\bullet$] The companion candidates must share the same proper motion as the M and L subdwarfs in our sample in each direction within 3$\sigma$: $\mu_{Obj}-3\sigma_{\mu.Obj}\leq \mu_{Comp} \leq \mu_{Obj}+3\sigma_{\mu.Obj}$.
\item [$\bullet$] The companion candidates must share the same distance as the object in our sample within 3$\sigma$: $d_{Obj}-3\sigma_{d.Obj}\leq d_{Comp} \leq d_{Obj}+3\sigma_{d.Obj}$. Here we add a restriction into the search to avoid too many spurious candidates, restricting the possible candidates to those with a maximum relative error in their distances of 20\%.
\item [$\bullet$] The companion candidates must lie within the search radius previously defined for every source in our sample. The search radius used for each subdwarf in our sample is presented in the last column of Table~A-2.
\end{itemize}
\section{Analysis}
\label{wide_sdM:analysis}
\subsection{Performed search}
\label{sec:search}
We looked for common proper motion companions to the 219 M and L subdwarfs in our sample with a search radius defined for each of them varying from 10.8\,arcmin to 9.4\,deg, corresponding to an interval of projected physical separations of 1.5 to 3.7\,pc. Despite the wide range of radii, the typical search radius is 1\,deg with 90\% of the sources covered within 2.28\,deg.
The lower limit of detection is given by the angular resolution of $Gaia$ DR2, and is 0.4\,arcsec. The detection of pairs with $Gaia$ DR2 is complete beyond 2.2\,arcsec \citep{gaiacollaboration18}, which is our adopted lower limit for completeness. This translates to minimum projected physical separations of between 20 and 1\,260\,au in our sample.

\subsection{Companion candidates}
\label{wide_sdM:companion_candidates}
We find 62 companion candidates around 12 M and L subdwarfs in our sample.

Table~\ref{tab_wide_sdM:number_wide_comp} shows the numerical identifiers of the sources with companion candidates, their coordinates, spectral types, search radii, and number of candidates.

We checked the re-normalised unit weight error (RUWE) in the $Gaia$ catalogue. This is an astrometric quality parameter that is high when the source has poor astrometric solutions \citep{lindegren18a}, and is sometimes influenced by the presence of another source. Hence, it can be used as an indication as to whether or not there could be an additional close companion. The additional documentation of $Gaia$ DR2 \citep{lindegren18b} shows that a value under 1.4 generally indicates a good solution because approximately 70\% of the sources have such a value. We noticed that 9 of our 62 candidates present RUWE values of between 1.4 and 6.6\@. Given that $Gaia$ DR2 exhibits RUWE values higher than 40, we do not consider 6.6 as a large value (i.e.\ meaning a poor astrometric solution). Therefore, we keep all 62 candidates for subsequent analysis. Table~A-3 lists the numerical identifier of each candidate companion together with their $Gaia$ DR2 source identifier, coordinates, proper motions, distances, and angular separations. As is the case for the rest of the tables contained in the Appendices, these data can be found in \textit{VizieR}.

\begin{table}
 \centering
 \caption[]{Number of companion candidates identified in this work.}
 \scalebox{0.86}[0.86]{
 \begin{tabular}{c@{\hspace{2mm}}c@{\hspace{2mm}}c@{\hspace{2mm}}c@{\hspace{2mm}}c@{\hspace{2mm}}c@{\hspace{2mm}}c}
 \hline
 \noalign{\smallskip}
 Id & RA (J2000) & DEC (J2000) & SpT  & \textit{J} & Radius & Num. \\
 & hh:mm:ss.ss & dd:mm:ss.s &  & mag & deg &  \\
 \noalign{\smallskip}
 \hline
 \hline
 \noalign{\smallskip}
11  & 01:18:24.90 & $+$03:41:30.4 & sdL0.0 & 18.2 & 0.50 & 32 \\
25  & 04:52:45.69 & $-$36:08:41.3 & esdL0.0 & 16.3 & 0.61 & 1 \\
73  & 10:46:57.93 & $-$01:37:46.4 & dM4.5/sdM5.0 & 16.5 & 0.33 & 1 \\
89  & 11:19:29.20 & $+$67:21:04.1 & sdM5.0$-$5.5 & 16.8 & 0.60 & 3 \\ 
107 & 12:41:04.75 & $-$00:05:31.6 & sdL0.0 & 18.5 & 0.45 & 2 \\
126 & 13:07:10.22 & $+$15:11:03.5 & sdL8.0 & 18.1 & 1.80 & 6 \\
128 & 13:09:59.60 & $+$05:29:38.7 & sdM6.5 & 18.4 & 0.22 & 1 \\
149 & 13:53:59.58 & $+$01:18:56.8 & sdL0.0 & 17.4 & 0.73 & 1 \\
150 & 13:55:28.24 & $+$06:51:14.6 & sdM5.5 & 17.6 & 0.39 & 1 \\
190 & 15:46:38.34 & $-$01:12:13.1 & sdL3.0 & 17.5 & 1.44 & 1 \\
213 & 22:59:02.15 & $+$11:56:02.1 & sdL0.0 & 17.0 & 0.84 & 1 \\
215 & 23:04:43.31 & $+$09:34:24.0 & sdL0.0 & 17.2 & 0.78 & 12 \\
 \noalign{\smallskip}
 \hline
 \end{tabular}
}
 \label{tab_wide_sdM:number_wide_comp}
\end{table}

We assess the validity of the candidates through visual inspection of the proper motion diagrams (PMDs), colour--magnitude diagrams (CMDs), and tangential velocity--distance diagrams. When possible, we also use radial velocities from the literature to compute galactocentric velocities, which serve us as membership indicators of the different Galactic populations, as explained below. Table~\ref{tab_wide_sdM:proposed_candidates} summarises whether each candidate agrees (\textit{Yes}), disagrees (\textit{No}), or is not conclusive (\textit{?}) with the position of the subdwarfs in the diagrams or whether it exhibits thick disc or halo kinematics. The last column of the table indicates whether the candidates are likely (\textit{Yes}) or doubtful (\textit{Yes?}) companions, or have been rejected (\textit{No}) as bound companions.

We analysed two optical and one infrared CMDs using the $Gaia$ $G$ and $RP$ passbands, the $i,z$ filters from SDSS, and the $J,K$ filters from 2MASS when available, or UKIDSS otherwise (when the 2MASS quality flags differ from `A' or `B'). For each CMD, we used the \textit{BT-Settl}\footnote{\url{https://phoenix.ens-lyon.fr/Grids/BT-Settl/CIFIST2011/ISOCHRONES/}} isochrones with metallicities [M/H] of $-$2.0 dex, $-$0.5 dex, and solar for comparison \citep{allard14} . 

\begin{figure*}[h]
  \caption*{\textbf{Id 150}}
  \vspace{5mm}
  \centering
  \includegraphics[width=0.6\linewidth, angle=0]{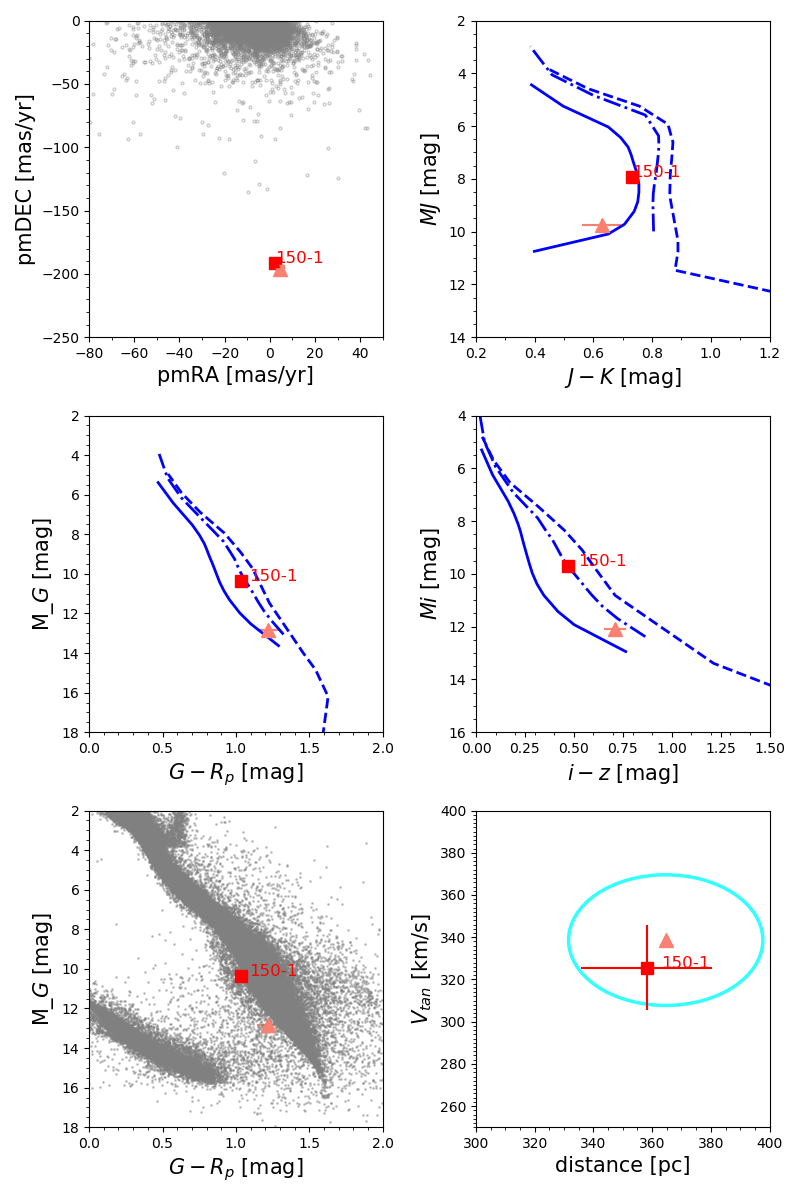}
  \vspace{5mm}
  \caption{PMD (top left panel), CMDs (top right and mid panels), HRD (bottom left panel), and tangential velocity--distance diagram (bottom right) for the target Id 150 and its candidate companion. The filled pink triangle represents the source under study, and the numbered red square represents the companion candidate. Grey dots in the PMD represent field stars, and in the HRD they are $Gaia$ DR2 sources with parallaxes larger than 10\,mas used as a reference. The blue solid, dashed, and dotted lines stand for the [M/H]\,=\,$-$2.0, the [M/H]\,=\,$-$0.5, and the [M/H]\,=\,0.0 \textit{BT-Settl} isochrones in the CMDs, respectively. The blue dotted line in the tangential velocity plot (not visible here) marks the value $V_{tan}=36\,km\,s^{-1}$ which is the mean value for field stars \citep{zhang18a}, and the light blue ellipse around Id 150 indicates its values of $V_{tan}\pm \sigma$ and $d \pm \sigma$.}
  \label{fig_wide_sdM:plot_CMD_PM_150}
\end{figure*}

We used the 10\,Gyr isochrones because ultracool subdwarfs are old objects. We complemented the $Gaia$ DR2 CMD with the observational HRD in the same bands, plotting $Gaia$ sources with parallaxes larger than 10\,mas as a reference. All diagrams used for the analysis are displayed in Appendix~\ref{wide_sdM:appendix_figures_CMD}, except from Id 150, which is shown as an example in Fig.~\ref{fig_wide_sdM:plot_CMD_PM_150}.

Radial velocities of each component of a physically bound system should be similar because of their presumably common origin. Therefore, we looked for radial velocities of every subdwarf with candidate companions using the SVO Discovery tool\footnote{\url{http://sdc.cab.inta-csic.es/SVODiscoveryTool/jsp/searchform.jsp}} developed and maintained by the Spanish Virtual Observatory, which performs a search through the \textit{VizieR} VO service.

None of the 12 subdwarfs in Table~\ref{tab_wide_sdM:number_wide_comp} have radial velocity measurements in the literature and therefore they could not be used for comparison. Nevertheless, we found values of radial velocities for 18 companion candidates \citep{gaiacollaboration18, zhong19, anguiano17, kervella19}, and with these data we are able to compute their $UVW$ Galactic space velocities from their coordinates, proper motions, and distances \citep{johnson87}. 
We can assign the companions to the different populations in the Galaxy (thin and thick discs, transition between thin and thick discs, and halo) according to the increasing galactocentric velocities of these populations towards the Galactic halo \citep{Bensby03,Bensby05}. This procedure is described in \citet{montes01a} and the updated version of the code used in this work will be published in Cort\'es-Contreras et al.\ (in prep). As stated before, subdwarfs belong to the old Galactic population, and therefore any candidate companion showing thick disc or halo kinematics would be a suitable companion. 

For all the subdwarfs and candidates, we can also use the tangencial velocities to assess old population kinematics, as in \citet{zhang18a}. 

\begin{itemize}

 \item [$\bullet$] Id 11: This subdwarf has a sdL0.0 spectral type and is located at about 176\,pc. We detected 32 possible companion candidates separated by between 1.7 and 29.9\,arcmin. The large number of spurious candidates is due to the high uncertainty in the proper motion computed in this work. As we are not able to reliably keep or discard any candidates, we reject all candidates above 1$\sigma$ in proper motion, leaving two companion candidates. Candidate Id number 17 shows higher proper motion than the subdwarf, lies between the $-$2.0 dex and $-$0.5 dex isochrones in the infrared CMD, and slightly above the solar-metallicity isochrone in the CMD with SDSS filters. Its position in the CMD with $Gaia$ photometry is slightly lower than the solar-metallicity isochrone, consistent with its position in the lower edge of the main sequence in the HRD. Moreover, its tangential velocity, which is higher than that of the field population, and within 1$\sigma$ of the velocity of the subdwarf, makes this source a potential companion candidate. The position of candidate Id number 19 in the HRD is above the solar-metallicity isochrone and within the main sequence, and its low tangential velocity and radial velocity suggest a location for this source of inside the thin disc; we therefore reject candidate number 19\@.
 \item [$\bullet$] Id 25: This subdwarf is an esdL0.0 source located at about 140\,pc with just one candidate separated by 1.9\,arcmin. It is interesting to remark that the companion candidate we find (denominated 25-1 hereafter) is Id 24 of our sample. The search program was not able to find a companion to Id 24 because of the very small margins of error on its proper motion provided by $Gaia$ DR2, despite the proximity of the values of both elements in the possible pair as shown in the PMD. The CMDs also show aligned positions for the subdwarf and for the candidate companion, except for the CMD with photometry from SDSS, which is not provided. The positions in the HRD are below the main sequence, and their tangential velocities are similar in terms of distance. We fully support this pair as likely companions.
 \item [$\bullet$] Id 73: This dM4.5/sdM5.0 source located at about 572\,pc has one companion candidate, 15.9 arcmin away. The PMD shows similar proper motions. We favour the UKIDSS photometry for this potential companion. This candidate lies at the same distance to the isochrones in all CMDs. From the positions of each component in the HRD, we suggest that both sources are probably solar metallicity rather than metal-poor, with a spectral type of dM2.0--dM2.5 as the spectroscopic follow-up suggests (section \ref{wide_sdM:spectro_ALFOSC}). Moreover, their tangential velocities show similarities in terms of their distance. All in all, we suggest both objects form a bound solar-metallicity pair. 
 \item [$\bullet$] Id 89: This is an sdM5.0--5.5 \citep{lodieu17a} object located at about 274\,pc with three companion candidates separated by 18.4, 34.7, and 40\,arcmin. The PMD shows very small values for the proper motions in all of the candidates and the subdwarf, and their positions in the CMDs are aligned. The $J-K$ colour of the ultracool subdwarf exhibits large error bars due to the poor quality flags in 2MASS photometry (Qflag\,=\,"CCD"). Nonetheless, this subdwarf lies below the main sequence in the HRD, as expected. On the contrary, the three companion candidates lie within or slightly above the main sequence, perhaps reflecting different metallicities. The tangential velocity diagram shows that none of the candidates are valid. However, all tangential velocities are below the mean value for field stars, suggesting that these objects do not have thick disk or halo kinematics. Consequently, we do not support companionship of any of them.
 \item [$\bullet$] Id 107: This source, with spectral type sdL0.0, located at 197\,pc has two companion candidates separated by 3.5 and 10.9\,arcmin. According to the PMD, candidate number 2 has proper motions that are not agreement with those of the subdwarf but still remain within $3\sigma$. Because of the faintness of the subdwarf, even the UKIDSS photometry suffers from large error bars in the $J-K$ colour diagram. For the candidates, the photometry comes from 2MASS. We reject candidate number 2 as a bound companion because of its position in the CMDs. The position of candidate number 1 in the HRD, below the main sequence, suggests subsolar metallicity, similar to our source. The subdwarf is not in $Gaia$ DR2 and therefore we cannot plot it on the HRD, but we are able to plot the positions of the companion candidate number 1, compatible with a low-metallicity source. Finally, its tangential velocity is within the range of halo objects. Therefore, we consider this pair physically associated.
 \item [$\bullet$] Id 126: This sdL8.0 source is located at 49\,pc. We identify six possible companion candidates with separations of between 34 and 98.5\,arcmin. The large uncertainty in the proper motion of the subdwarf prevents us from obtaining any reliable candidate from the PMD. Additionally, none of the candidates show photometric criteria consistent with metal-poor isochrones in the CMD and HRD\@. In particular, the radial velocity of candidate number 2 places it in the Galactic thin disc. In conclusion, we reject all of these candidates.
 \item [$\bullet$] Id 128: This subdwarf is a sdM6.5 source located at 547\,pc, with a single companion candidate at 8.5 arcmin. According to the PMD, the candidate has a similar motion to the subdwarf. However, their positions in the CMDs suggest inconsistent metallicities, corroborated by their location in the HRD where the companion follows with the main sequence solar-metallicity track. Additionally, their tangential velocities differ. Therefore, we discard the system. 
 \item [$\bullet$] Id 149: This source has a sdL0.0 spectral type and is located at 121\,pc. We detect a companion candidate at 36.1\,arcmin. The PMD shows that both objects have very similar proper motions. There is no photometry available in $Gaia$ for our subdwarf and the SDSS $i,z$ photometry of the candidate is clearly saturated, and so these CMDs do not provide useful information. Both objects show similar positions in the other CMD\@. The companion candidate is HD\,120981, a G2/3V star \citep{houk88} with metallicity in the range 0.09--0.27 \citep{ammons06,stevens17}. Combining tangential velocity and radial velocity, we find that the companion shows galactocentric velocities typical of the thin disk. We discard this system as a physical pair.
 \item [$\bullet$] Id 150: This source has a sdM5.5 spectral type and is located at about 365\,pc. It has just one companion candidate separated by 3.8\,arcsec, which we visually confirm as a common proper motion pair in \textit{Aladin}. The positions in the PMD and CMDs support their companionship. Both stars are under the main sequence in the HRD, which supports their metal-poor nature. The tangential velocities of both sources are very similar and well above the mean value of field sources of 36\,km\,s$^{-1}$ \citep{zhang18a}, in agreement with an old population member. The companion is in the LSPM catalogue \citep{lepine05d} with the identifier J1355$+$0651\@. We propose that this system is a bound pair.
 \item [$\bullet$] Id 190: This is a sdL3.0 source at $\sim$61\,pc with one candidate companion at 8.9\,arcmin. We do not have enough information to discard or confirm the candidate. The PMD shows very similar proper motions and the positions in one of the CMDs agree. The tangential velocity is above the mean value of field sources, outside the error limits of the value of the subdwarf, but with very close values. The current information is not conclusive but we decided to include this possible pair in a subsequent analysis.
 \item [$\bullet$] Id 213: This is a sdL0.0 located at about 105\,pc with a single candidate companion at 36.8\,arcmin. The large uncertainties in the input parameters of our subdwarf provide a unique candidate companion with very different proper motions. Its position on the CMDs and HRD suggest a likely pair, but its tangential velocity diagram does not, and the kinematic analysis using its radial velocity suggests the companion as a thin disk star. We reject this system as a probable pair.
 \item [$\bullet$] Id 215: This subdwarf is a source with sdL0.0 spectral type located at 114\,pc with 12 candidates separated by 22.4 to 46.3\,arcmin. The large error bars in the proper motions of the subdwarf lead to a large number of spurious candidates. As in the case of Id 11, we refute all candidates with proper motions above 1$\sigma$ of the proper motion of the subdwarf. Only candidate numbers 2, 8, and 9 remain. There is no available photometry for the subdwarf in $Gaia$, and these three candidates are aligned in the rest of the CMDs, except number 2 which is not aligned in the optical CMD. Based on the HRD, we infer that the metallicity classes of candidates 2 and 9 probably differ from the subdwarf. Using the radial velocity of candidate 8, we place it in the thin disk. The tangential velocity of the three candidates is lower than the mean velocity of field stars. Consequently, we reject all of them.
\end{itemize}

\begin{table*}
 \centering
 \caption[]{\small{Logs of spectroscopic observations of three companion candidates}
}
 \scalebox{0.85}[0.85]{
 \begin{tabular}{@{\hspace{0mm}}l c c c c c c c c c@{\hspace{0mm}}}
 \hline
 \noalign{\smallskip}
Name  & RA (J2000)  & DEC (J2000) &  Instr & $i'$  & Date     & UT-MID       & Airm & ExpT    & SpT \cr

      & hh:mm:ss.ss   & dd:mm:ss.s  &  & mag   & DDMMYYYY & hh:mm:ss.ss  & & sec  & \cr
 \hline 
 \hline 
 \noalign{\smallskip}
Id 150-1  & 13:55:28.38 & $+$06:51:18.5 & ACAM & 17.46  & 02022019 & 06:15:03.475 & 1.08 &  900   & sdM1.5$\pm$0.5 \cr
Id 73-1  & 10:47:13.20 & $-$01:22:21.4 & ALFOSC & 17.60  & 13122020 & 05:15:07.02 & 1.22 &  2100   & dM2.0--dM2.5 \cr
Id 107-1  & 12:40:35.18 & $-$00:13:28.7 & ALFOSC & 16.71  & 13122020 & 06:00:48.00 & 1.41 &  900   & dM/sdM5.0$\pm$0.5 \cr
 \noalign{\smallskip}
 \hline
 \label{tab_wide_sdM:log_obs_spectro}
 \end{tabular}
 }
\end{table*}

\begin{figure}
  \centering
  \includegraphics[width=\linewidth, angle=0]{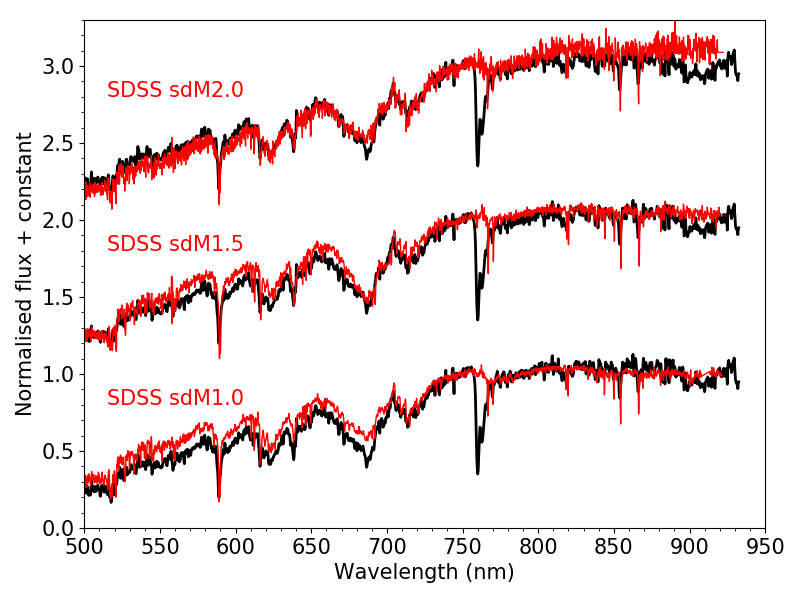}
  \includegraphics[width=\linewidth, angle=0]{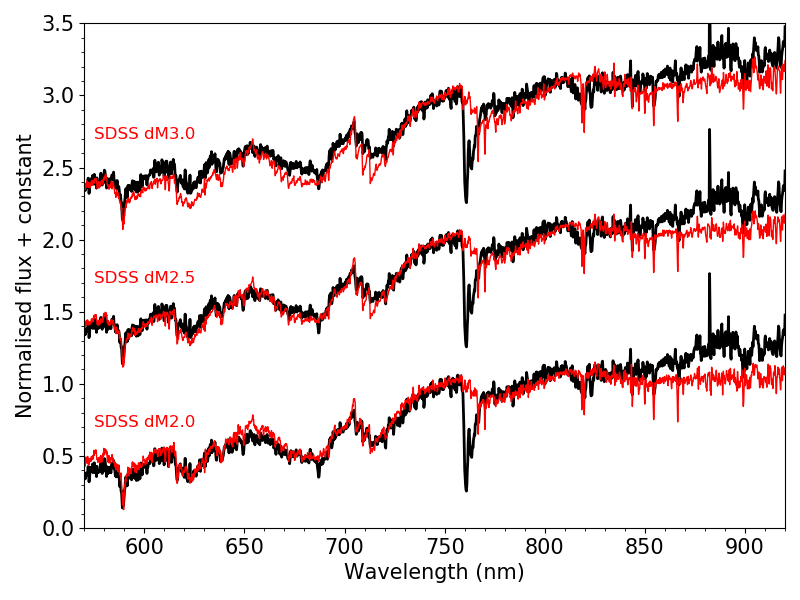}
  \includegraphics[width=\linewidth, angle=0]{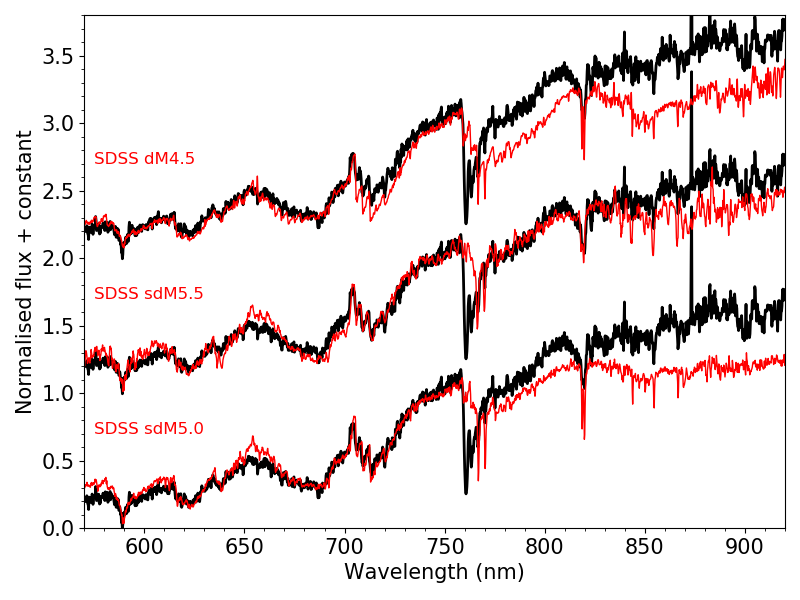}
   \caption{\small{WHT/ACAM optical spectrum (black) of the companion of subdwarf Id 150 (top)
   and NOT/ALFOSC spectra (black) of Id 73 (middle) and Id 107 (bottom) compared to Sloan 
   solar-type and subdwarf templates (red).}
}
   \label{fig_wide_sdM:plot_spectra_WHT_ACAM}
\end{figure}

\subsection{Spectroscopic follow-up}
\label{wide_sdM:spectro}
\subsection{WHT ACAM optical spectroscopy}
\label{wide_sdM:spectro_ACAM}

We collected a low-resolution optical spectrum of the candidate companion Id 150-1 (i.e.\ the primary) with the auxiliary-port camera (ACAM) mounted on the Cassegrain focus of the 4.2-m William Herschel telescope at Roque de los Muchachos observatory in La Palma (Table~\ref{tab_wide_sdM:log_obs_spectro}). We carried out the observations in service mode on the night of 1 February 2019\@. The night of 1 February 2019 was clear with variable seeing between 1.0 and 1.6\,arcsec after UT\,$\sim$\,3h when the object was observed. 

ACAM is permanently mounted on the telescope as an optical imager and spectrograph. We used the VPH grism with a slit of 1.0\,arcsec to cover the 350--940\,nm wavelength range at a resolution of 430 and 570 at 565\,nm and 750\,nm, respectively. We did not use a second-order blocking filter, resulting in light contamination beyond 660\,nm{}. We use single on-source integrations of 900\,s (Table~\ref{tab_wide_sdM:log_obs_spectro}).

We reduced the data in a standard manner under {\tt IRAF} \citep{tody86,tody93}. 
We subtracted a median-combined bias frame to the target's frame and later divided by the median flat-field. We extracted the one-dimensional spectrum in an optimal way by selecting the size of the aperture and the background regions on the left and right side of the target. We calibrated the spectra in wavelength with CuNe arc lamps taken just after the target, yielding rms better than 0.3\,nm. Finally, we calibrated our targets in flux with a spectro-photometric standard observed with the same setup during the night \citep[Ross640; DAZ5.5][]{greenstein67,monet03}. However, the flux calibration is uncertain beyond 660\,nm because the second-order blocking filter was not in place. We display the spectrum in Fig.~\ref{fig_wide_sdM:plot_spectra_WHT_ACAM} along with Sloan metal-poor templates \citep{savcheva14}. From the spectral fits, we derive a spectral type of sdM1.5$\pm$0.5\@.

\subsection{NOT ALFOSC optical spectroscopy}
\label{wide_sdM:spectro_ALFOSC}
We obtained low-resolution optical spectra of two additional candidate companions (Id 73-1 and Id 107-1) with the Alhambra Faint Object Spectrograph and Camera (ALFOSC) on the 2.5-m Nordic Optical telescope (NOT) at Roque de los Muchachos observatory in La Palma (Table~\ref{tab_wide_sdM:log_obs_spectro}). We collected the spectra as part of a service programme (SST2020-506; PI Cort\'es-Contreras) on the night of 12 December 2020, which presented cirrus and seeing between 1 and 1.5 arcsec during the observations between UT\,=\,5h--6h.

ALFOSC is equipped with a 2048$\times$2048 CCD231-42-g-F61 back-illuminated, deep-depletion detector sensitive to optical wavelengths. The pixel size is 0.2138 arcsec and the field of view is 6.4$\times$6.4 arcmin across. We employed the VPH grism \#20 and a slit of 1.3 arcsec covering the 565--1015\,nm at a resolution of about 500\@. We use an on-source integration of 2100\,s and 900\,s for Id 73-1 and Id 107-1, respectively (Table~\ref{tab_wide_sdM:log_obs_spectro}). A spectro-photometric standard star, Feige 66 \citep[sdB1; $V$\,=\,10.59 mag;][]{berger63} was observed to characterise the response of the detector.

We reduced the data with IRAF following standard procedures. We combined the bias and flat frames before subtracting each science spectrum by the median-combined bias and subtracting the normalised flat field. We optimally extracted the 2D spectra choosing the aperture and background regions interactively. We calibrated the spectra in wavelength with ThAr lamps taken at the end of the night. We corrected the science spectra with the response function derived from Feige 66\@. The NOT ALFOSC optical spectra, normalised at 750\,nm, are displayed in Fig.~\ref{fig_wide_sdM:plot_spectra_WHT_ACAM}
along with Sloan spectral M-type templates at solar and subsolar metallicities \citep{bochanski07a,savcheva14}. From the direct comparison with Sloan spectral templates, we classify Id 73-1 as a solar-metallicity M2.0--M2.5 dwarf. For Id 107-1, we find that this source shows features intermediate between a solar-metallicity and subdwarf with a mid-M spectral type, and we adopt a classification of dM/sdM5.0$\pm$0.5\@.

\section{Results of the search}
\label{wide_sdM:results_search}
Following the search and analysis performed in previous sections, we identified six M and L subdwarfs with potential candidate companions, some of them with high probabilities.

To further characterise these stars and their candidates companions, we built their spectral energy distribution (SED) using the VO SED Analyzer \citep[VOSA;][]{bayo08}, which provides estimates of stellar physical parameters from the SED fitting to different collections of theoretical models. In this work, we used the \textit{BT-Settl} theoretical models \citep{baraffe15}. We limited the range of gravities to $4\leq log(g)\leq 6$, which are the normal values for old dwarfs \citep{cifuentes20}. We prove that metallicity does not have a significant impact on the determination of effective temperatures in VOSA, because it takes values in the range of $\pm$100\,K when metallicity varies from -3 to 0\,dex. Finally, we took the effective temperature proposed as the best one by VOSA with an error of $\pm$100\,K ---instead of $\pm$50\,K which would correspond to the grid error established by VOSA--- to provide a more realistic margin. The obtained effective temperatures of the candidate companions will give us information on their spectral types. 

In the calculation of the masses, we used different methods to estimate them. In the case of sdL sources, we took the values of Table\,3 and Figure\,5 from \cite{zhang18a} in the range of 0.08--0.09\,$M_{\astrosun}$. For sdM sources, we estimate a range of masses with the isochrone model\footnote{\url{http://perso.ens-lyon.fr/isabelle.baraffe/BCAH97_models}} from \citet{baraffe97}, using as input the effective temperature obtained by VOSA and the J magnitude from the UKIDSS catalogue (or 2MASS if not available in  UKIDSS ). For solar metallicity sources, we used the isochrone model\footnote{\url{http://perso.ens-lyon.fr/isabelle.baraffe/BHAC15dir/}} from \citet{baraffe15}.

In the estimation of the spectral type, for the case of low-metallicity dwarfs, we check Table\,2 from \citet{lodieu19b}, and Figure\,4 from \citet{zhang18a}. For solar metallicities, we can estimate the spectral type following \citet{reid05b}.
Additionally, we intensively browsed the literature for any additional relevant information related to the companion candidates (spectral type, metallicity, age), including references to any known physically bound or unrelated companions.

\begin{figure}
  \centering
  \includegraphics[width=0.45\textwidth]{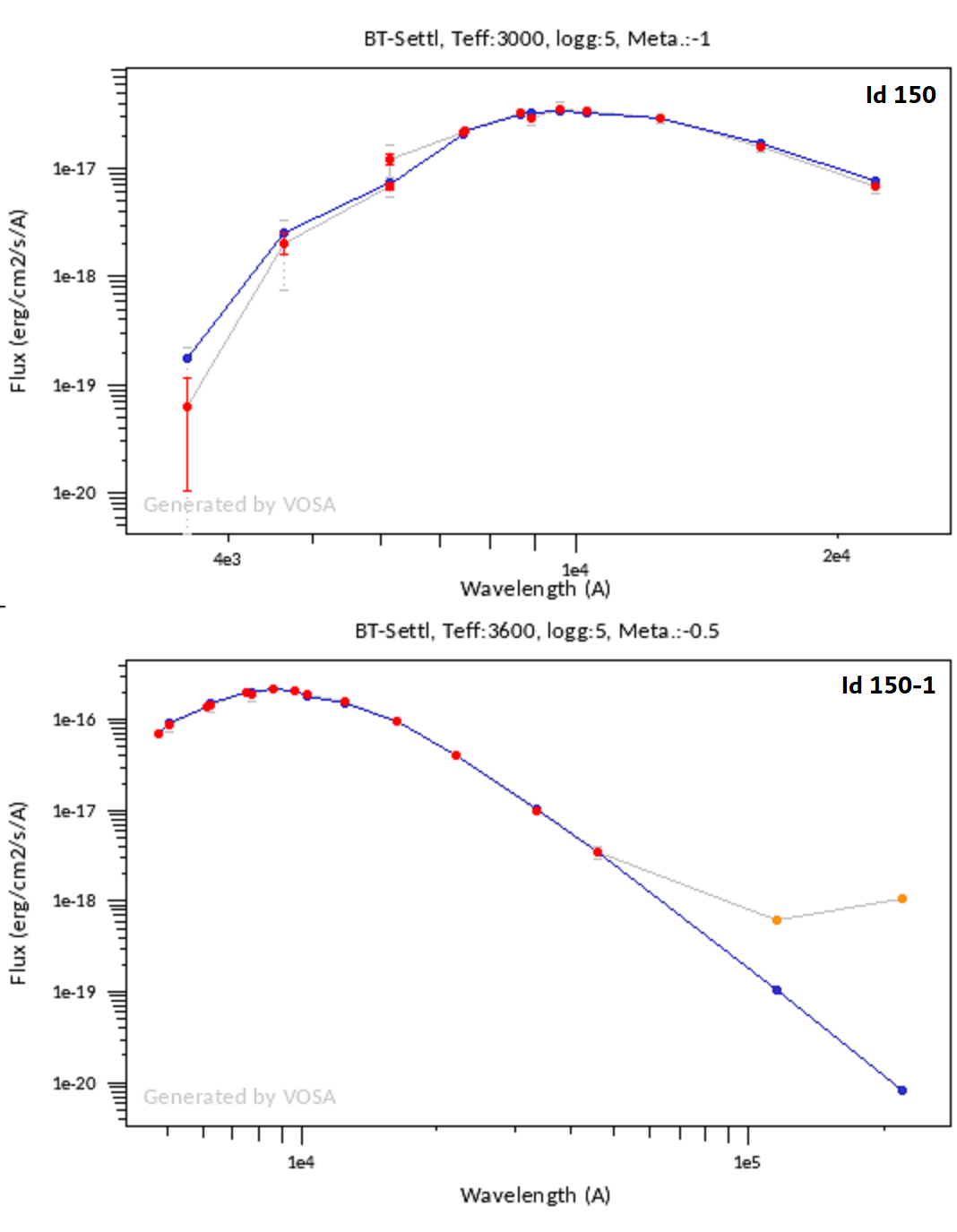}
   \caption{VOSA diagrams for the source Id 150  (top) and its candidate companion (bottom). The shown elements are: Observed flux + 3-sigma points (pale grey line + error bars), fitted flux (red dots + error bars), no fitted points (orange dots + error bars), and \textit{BT-Settl} model (blue line).}
   \label{fig:vosa150}
\end{figure}

\begin{itemize}
    \item [$\bullet$] Id 11: This subdwarf has a spectral type of sdL0.0 with one remaining candidate companion (Id 11-17). We obtained their SEDs with VOSA, from which we derived effective temperatures of 2\,600$\pm$100\,K and 2\,900$\pm$100\,K for Id 11 and Id 11-17, respectively. Using \citet{lodieu19b} and \citet{zhang18a}, we estimated its spectral type as sdM8.0$\pm$0.5. We also calculated its binding energy, which is very low (about ten times lower than the accepted minimum energy to be considered a bound pair). We considered the possibility of both objects being part of a multiple system, but the small RUWE value of the candidate (1.11) points towards a single star and therefore a low chance of such a multiple system having a larger binding energy.
    \item [$\bullet$] Id 25: This subdwarf and its candidate companion are extreme subdwarfs, and are reported as a known pair by \citet{zhang19g}. Their spectral types are esdL0.0 and esdM1.0 respectively, which we support from the temperatures derived by VOSA (2\,800$\pm$100\,K and 3\,700$\pm$100\,K respectively). All the obtained data are in line with the reported ones by \citet{zhang19g}, with small differences due to the fact that Zhang estimated the properties with a lower metallicity than that used here (he used $[M/H]=-1.4$\,dex). The binding energy of the system is consistent with a physically bound pair.
    \item [$\bullet$] Id 73: This object in our sample is a solar metallicity dwarf with spectral type dM4.5, and an effective temperature of 3\,100$\pm$100\,K derived from its SED. The possible companion with Id 73-1 is a solar metallicity dwarf with a temperature of 3\,500$\pm$100\,K as NOT ALFOSC optical spectrum suggests, corresponding to a spectral type of dM2.0--2.5. With this new spectral type we recalculated the distance of the companion candidate through spectroscopic methods \citep{zhang13} and obtained 607.2$\pm$98.2\,pc, a value that is closer to the distance of Id 73 (instead of 860.9$\pm$159.7\,pc, the value obtained from the parallax provided by \textit{Gaia} DR2, as shown in Table~A-3). This pair has a low binding energy, about 70\% of the minimum to be a bound pair. The RUWE value of the brightest source of the pair is low (0.95), indicating that it is unlikely to be part of a multiple system. The analysis of the spectrum at our resolution shows all lines to be single. At this stage, the pair seems to be bound.
    \item [$\bullet$] Id 107: The subdwarf has a spectral type of sdL0.0, and we suggest that its candidate companion is also metal poor in light of its position on the HRD. There are not enough photometric points to build the SED of the subdwarf, but we estimate its temperature to about 2\,600$\pm$100\,K from its spectral type using Figure~4 of \citet{zhang18a}. VOSA provides the value of 3\,300$\pm$100\,K for the companion candidate, for which we estimate a spectral type of dM/sdM5.0$\pm0.5$, in agreement with the spectral classification of the NOT ALFOSC spectrum. The optical spectrum of the wide companion suggests that this system might have an intermediate metallicity between solar and $-$0.5 dex (typical of subdwarfs), and its lines do not appear deblended. In light of the possibility that this companion source is a dwarf, we calculated its range of mass using isochrone models from \citet{baraffe97} and \citet{baraffe15}. Here, the value of the binding energy is also low, three times lower than the minimum, and the companion candidate has a RUWE of 0.98. We conclude that the pair seems to be a bound system.
    \item [$\bullet$] Id 150: This subdwarf has a spectral type of sdM5.5, and the companion candidate is also a subdwarf, as suggested from its HRD. The effective temperatures derived by VOSA (see Figure~\ref{fig:vosa150}) are 3\,000$\pm$100\,K for the subdwarf and 3\,600$\pm$100\,K for the companion candidate. The temperature of the companion is in agreement with the spectral type of sdM2.0$\pm$0.5 from our spectroscopic follow-up (Section~\ref{wide_sdM:spectro}) based on Table~2 of \citet{lodieu19c}. The two sources are very close (about 1\,360\,au), and we visually confirm them to be a comoving pair in \textit{Aladin}. Their binding energy is more than 40 times the minimum energy of gravitationally bound pairs, reinforcing the system as a true pair. We conclude that this pair is the most secure in our sample.
    \item [$\bullet$] Id 190: The subdwarf Id 190 has a spectral type of sdL3.0. The effective temperatures provided by VOSA for the subdwarf and the companion candidate are 2\,200$\pm$100\,K (consistent with the spectral type) and 3\,000$\pm$100\,K, respectively. As we did for other subdwarfs, we estimated the spectral type of the companion to sdM7.0$\pm$0.5 from its temperature \citep{lodieu19b}. Their masses and distances provide us a value for the binding energy of 88\% of the minimum; with a low RUWE of 1.17 for the companion candidate, we conclude that this pair has a low chance of being a real system. We conclude that this is the most doubtful pair in our sample.
    
    \end{itemize}
    
From the previous analysis, we propose one confirmed binary system (Id 150), four likely pairs (Id 11, Id 25, Id 73, and Id 107), and a doubtful pair (Id 190). Their identifiers, names, spectral types, effective temperatures, masses, proper motions, distances, physical and projected angular separations, and binding energies are provided in Table~\ref{tab_wide_sdM:proposed_pairs}. 

As mentioned above, the most commonly accepted value of the minimum gravitational binding energy is 10$^{33}$\,J\@. In our case, only two pairs with the closest separations, Id 25/25-1 (0.08\,pc) and Id 150/150-1 (0.007\,pc), fulfil this criterion. The other four systems have much lower gravitational binding energies. However, this study does not account for hidden mass in the systems in the form of spectroscopic binaries or fainter sources not detected with $Gaia$ DR2. In terms of projected physical separations, the widest pair is separated by 2.64\,pc (Id 73/73-1). The classical limit of separation for binaries is 0.1\,pc \citep{caballero09}, but there are studies that increase this number to 1.0\,pc \citep{caballero10}, or even to 1--8\,pc \citep{shaya11}. At those distances, the pairs are less likely bound for an extended lifetime, and the sources of our sample are very old. This is not the case for our proposed subdwarfs, whose separation from the companions are lower than 0.7\,pc. In the case of the solar-type system Id 73, although it has a separation above the classical limits, the component sources are field dwarfs with a mean age of less than the metal-poor population.

\begin{table*}[h]
 \centering
 \caption{Physical parameters of the proposed systems.}
 \scalebox{0.75}[0.75]{
 \begin{tabular}{@{\hspace{2mm}}l@{\hspace{2mm}}l@{\hspace{2mm}}c@{\hspace{2mm}}c@{\hspace{2mm}}c@{\hspace{2mm}}c@{\hspace{2mm}}c@{\hspace{2mm}}c@{\hspace{2mm}}c@{\hspace{2mm}}c@{\hspace{2mm}}c@{\hspace{2mm}}c@{\hspace{2mm}}c@{\hspace{2mm}}}
\hline
\noalign{\smallskip}
 Id & Name & SpT & [M/H]\tablefootmark{a} & T$_{eff}$ & Mass & J & $\mu_{\alpha}\cos\delta$ & $\mu_\delta$ & Distance & \multicolumn{2}{c}{Separation} & W\\
    &      &     & dex &     K   & M$_{\astrosun}$ & mag & mas\,a$^{-1}$ & mas\,a$^{-1}$ & pc & pc & arcmin  & $10^{33}$\,J  \\
\hline
\hline
\noalign{\smallskip}
\multicolumn{13}{c}{\textit{Confirmed pairs}} \\
\hline
\noalign{\smallskip}
150   & Gaia DR2 3720832015084722304 & sdM5.5$\pm$0.5 & --0.5 & 3\,000$\pm$100 & 0.092--0.125  & 17.55$\pm$0.04 & 4.4$\pm$2.5 & $-$195.9$\pm$2.0 & 364.6$\pm$33.1 &  &  &  \\
150-1 & Gaia DR2 3720832010789680000 & sdM2.0$\pm$0.5 & --0.5 & 3\,600$\pm$100 & 0.149--0.279  & 15.71$\pm$0.01 & 2.5$\pm$0.4 & $-$191.7$\pm$0.3 & 358.2$\pm$22.2 &  0.007 & 3.8\tablefootmark{b} & 42.6 \\
\noalign{\smallskip}
\hline
\noalign{\smallskip}
\multicolumn{13}{c}{\textit{Likely pairs}} \\
\hline
\noalign{\smallskip}
11    & ULAS J011824.89$+$034130.4     & sdL0.0$\pm$0.5  & --0.5 & 2\,600$\pm$100 & 0.080--0.090  & 18.18$\pm$0.05 & 19.9$\pm$29.0 & $-$33.8$\pm$27.0  & 176.2$\pm$13.7 &  &  &  \\
11-17  & Gaia DR2 2562996857437494528 & sdM8.0$\pm$0.5  & --0.5 & 2\,900$\pm$100 & 0.089--0.135  & 15.72$\pm$0.06 & 41.2$\pm$1.2 & $-$45.1$\pm$0.6 & 177.1$\pm$17.7 & 0.69  & 13.4 & 0.10 \\
\noalign{\smallskip}
\hline
\noalign{\smallskip}
25    & 2MASS J04524567$-$3608412   & esdL0.0$\pm$0.5  & $-$1.0 & 2\,800$\pm$100 & 0.080--0.090  & 16.26$\pm$0.10 & 147.5$\pm$0.8 & $-$168.1$\pm$1.0  & 140.2$\pm$10.0 &  &  &  \\
25-1  & Gaia DR2 4818823636756117504 & esdM1.0$\pm$0.5  & $-$1.0 & 3\,700$\pm$100 &  0.178--0.275 & 13.77$\pm$0.03 & 148.5$\pm$0.1 & $-$168.7$\pm$0.1 & 137.3$\pm$0.7 & 0.08  & 1.92 &  2.65 \\
\noalign{\smallskip}
\hline
\noalign{\smallskip}
73    & SDSS J10465793$-$0137464       & dM4.5$\pm$0.5  & 0.0 & 3\,100$\pm$100 & 0.119--0.413 & 16.54$\pm$0.01 & $-$25.2$\pm$1.0 & $-$9.8$\pm$0.8  & 572.5$\pm$175.6 &  &  &  \\
73-1  & Gaia DR2 3802720750608315392 & dM2.0--2.5  & 0.0 & 3\,500$\pm$100 & 0.289--0.524 & 15.88$\pm$0.01 & $-$22.4$\pm$0.4 & $-$10.1$\pm$0.3 & 607.2$\pm$98.2 &  2.64 & 15.9 &  0.70 \\
\noalign{\smallskip}
\hline
\noalign{\smallskip}
107   & ULAS J124104.75$-$000531.4     & sdL0.0$\pm$0.5 & --0.5 & 2\,600$\pm$100 &  0.080--0.090  & 18.46$\pm$0.10 & $-$42.9$\pm$8.0 & $-$26.2$\pm$6.0 & 196.5$\pm$15.3 &  &  & \\
107-1 & Gaia DR2 3695978963488707072 & dM/sdM5.0$\pm$0.5 & --0.5-0.0 & 3\,300$\pm$100 & 0.107--0.289 & 14.73$\pm$0.04 & $-$36.9$\pm$0.3 & $-$20.7$\pm$0.1 & 177.8$\pm$3.7 &  0.62 & 10.9 & 0.36 \\
\noalign{\smallskip}
\hline
\noalign{\smallskip}
\multicolumn{13}{c}{\textit{Doubtful pairs}} \\
\hline
\noalign{\smallskip}
190   & ULAS J154638.34-011213.0 & sdL3.0$\pm$0.5 & --0.5 & 2\,200$\pm$100 & 0.080--0.090 & 17.51$\pm$0.04 & $-$49.9$\pm$8.6 &  $-$107.1$\pm$7.6 & 61.1$\pm$6.0 &  &  &  \\
190-1 & Gaia DR2 4404197733205321216 & sdM7.0$\pm$0.5 & --0.5 & 3\,000$\pm$100 & 0.092--0.183 & 12.85$\pm$0.02 & $-$52.8$\pm$0.2 & $-$122.7$\pm$0.1 & 69.8$\pm$0.4 & 0.16 & 8.9 & 0.88 \\
\noalign{\smallskip} 
\hline
 \end{tabular}
}
\label{tab_wide_sdM:proposed_pairs}
\tablefoot{\tablefoottext{a}{Not calculated but derived from the metallicity class definition.
Uncertainty of 0.5 dex.} \tablefoottext{b}{This value in arcsec.}} 

\end{table*}

\section{Discussion on multiplicity}
\label{wide_sdM:discuss}
%
\subsection{General considerations about multiplicity}
\label{wide_sdM:discuss_SpT}

The binary frequency decreases with decreasing spectral type \citep{fontanive18}. Over 70\% of massive B and A-type stars are part of binary or hierarchical systems \citep{kouwenhoven07b,peter12}. The incidence of multiplicity is about 50--60\% for solar-type stars \citep{duquennoy91,raghavan10}, and about 30--40\% for M dwarfs \citep{fischer92,delfosse04,janson12}. Later surveys found that the multiplicity fraction of M-dwarfs drops to 23.5--42\% \citep{ward15,cortes17b}. Thus, the overall trend is that the multiplicity rate of main sequence stars decreases with mass \citep{jao09a}.

The total multiplicity frequency od population II stars with spectral types from F6 to K5 (masses between $0.7M_{\astrosun}$ and $1.3M_{\astrosun}$) is 39$\pm$3\% \citep{jao09a}, dropping to 26$\pm$6\% for stars with spectral types between K6 and M7 \citep[0.1--0.6M$_{\astrosun}$;][]{rastegaev10}. Spectroscopic binaries with separations of less than a few astronomical units (au) among population II stars seem to be equally frequent to younger, higher metallicity stars \citep{stryker85,latham02,goldberg02a}. This fact appears to hold for separations of a few tens od au \citep{koehler00b,zinnecker04}, and at very wide separations for GKM stars \citep{allen00,zapatero04a}, suggesting that metallicity might not have a strong impact on the formation of wide double and multiple systems. At lower masses, the frequency of early-M subdwrafs appears significantly lower \citep[3.3$\pm$3.3\%][]{riaz08a,lodieu09c} than the multiplicity of solar-metallicity M dwarfs \citep[23.5--42\%;][]{ward15,cortes17b} over similar separation ranges.
\subsection{Multiplicity}
\label{wide_sdM:discuss_multiplicity}
Our search for companions in our sample of 219 low-metallicity dwarfs found one solar-metallicity M-M pair (Id 73/73--1), which was confirmed by analysis of its optical spectrum. We exclude this system from the subsequent discussion because we focus on metal-poor systems. For the final sample of 218 M5-L8 subdwarfs, our search revealed five candidate companions: one clear metal-poor M-M pair (Id 150/150--1), and four M-L pairs. The physical projected separations of the companion candidates lie between 3.8\,arcsec and 13.4\,arcmin, equivalent to 1\,360 (0.007\,pc) and 142\,400\,au (0.69\,pc). In all cases, the candidate companions are warmer than our subdwarfs. We infer a binary fraction of 2.29$\pm$2.01\% from the five metal-poor pairs among the 218 sources, assuming Poisson statistics and a Wald 95\% confidence interval (Wald interval\,$=(\lambda - 1.96\sqrt{\lambda/n}, \lambda + 1.96\sqrt{\lambda/n})$, where $\lambda$ is the number of successes in $n$ trials). In any case, we find one clear co-moving low-metallicity pair (Id 150), placing the minimum probability of finding a metal-poor M5--L8 dwarf in a binary system at 0.46\small $_{-0.46}^{+0.90}$\% \normalsize. We discuss the binary fractions as a function of spectral type and/or metallicity class below, and summarise our results in Table~\ref{tab_wide_sdM:binarity}.

We identify one M-M subdwarf pair, Id 150, with a separation of 1\,360 au. We have a total of 167 M5--M9.5 metal-poor dwarfs in our sample and found only one M-M pair, yielding a frequency of 0.60\small $_{-0.60}^{+1.17}$\% \normalsize . In terms of metallicity, Id 150 is a subdwarf, and in our sample we have 97 M subdwarfs (sdMs), yielding a binarity of 1.03\small $_{-1.03}^{+2.02}$\% \normalsize. Id 25 has a lower metallicity because it is an extreme M subdwarf system, and so the frequency of esdM systems is 1.89\small $_{-1.89}^{+3.70}$\% \normalsize (1/53), assuming Poisson statistics. We did not find any companion to the 19 usdM in our sample, implying an upper limit of 5.3\% on the binary fraction of ultra subdwarfs.

For the M-L pairs, the subdwarf of our sample is the L-type while the companion is the M-type. Following the general convention, the primary is the most massive of the pair, and hence the L subdwarfs are the secondaries. Therefore, we find three L-type secondaries around M subdwarfs (plus one doubtful) out of 49 L subdwarfs in our sample, yielding a binarity of 6.12\small $_{-6.12}^{+6.93}$\% \normalsize, that can increase to 8.16$\pm$8.00\% \normalsize if we include the doubtful pair (Table \ref{tab_wide_sdM:proposed_pairs}). This means that 6.12\% of the L subdwarfs of our sample are part of a multiple system with a maximal physical projected separation of 0.69\,pc (13.4\,arcmin) while our search is sensitive to separations up to 1.5\,pc. In this case, we are not strictly talking about a binary fraction because our search in \textit{Gaia} DR2 is not sensitive to lower mass companions to the L subdwarfs. 

We should mention that two sources in our sample have known companions in the WDS catalogue \citep{mason01}. On the one hand, Id\,3 (WDS\,00259$-$0748) has a companion at 2.1\,arcsec with a magnitude of 22.6 in the F775W optical filter \citep{riaz08a}. On the other hand, Id\,154 (WDS\,14164$+$1348, also known as SDSS J1416$+$13AB) is an sdL7 source with a T5 companion separated by 9.3\,arcsec with $J$\,=\,17.26\,mag \citep{scholz10a,schmidt10a,bowler10a,burningham10a}. Both companions are beyond the reach of our survey because of their faintness. We do not identify new more massive companions to both objects. 

For the same separation range, our frequency is much lower than the multiplicity of F6$-$K3 stars \citep[44$\pm$3\%;][]{fischer92} and K7$-$M6 \citep[23.5$\pm$3.2\%;][]{ward15}; see also Table~1 in \citet{cortes17b}. Our result is more in line with the frequency of early-M subdwarfs based on high-resolution Hubble Space Telescope and Lucky Imaging \citep{riaz08a,lodieu09c}. Our multiplicity is much lower than the frequency of GKM subdwarfs \citep[13--15\%;][]{zapatero04a}.

\citet{moe19} state that there is no difference between the frequencies of solar or subsolar metallicity samples among wide ($a$\,$>$\,1\,000\,au) binaries. Similarly, \citet{El_Badry19a} conclude that the wide binary fraction is almost constant with metallicity at large separations ($a\geq 250$\,au), but decreases quickly with metallicity at smaller separations. These statements do not seem to hold for ultracool subdwarfs with spectral types later than M5\@.

The binary frequency of M subdwarfs still remains unclear because of poor statistics. \citet{jao09a} studied a sample of 32 K and 37 M subdwarfs and derived a multiplicity of 26$\pm$6\% for separations larger than 110 au and 6\% for lower separations. To provide a more complete picture, we combined the samples of \citet{gizis00c}, \citet{zapatero04a}, \citet{riaz08a}, and \citet{lodieu09c} to perform a search for companions around metal-poor GKM dwarfs using CCD and Lucky Imaging. We selected all M subdwarfs in this compilation and built a new sample adding all M subdwarfs from our sample, regardless of their metallicity class. Only the work of \citet{zapatero04a} does not provide spectral types or effective temperatures and therefore we recovered M type stars from their colours using infrared and optical photometry and the updated version of Table~4 in \citet{pecaut13}\footnote{\url{https://www.pas.rochester.edu/~emamajek/EEM_dwarf_UBVIJHK_colors_Teff.txt}}. These latter authors searched for companions around subdwarfs with separations of between 0.1--0.2 and 25\,arcsec. As $Gaia$ is able to resolve pairs at 2.2\,arcsec, we neglect all binaries found at closer separations and do not consider any companion at separations above 25\,arcsec in order to obtain a coherent binary fraction. This new sample contains 279 objects with 264 low-metallicity M and L dwarfs (215 M, 49 L), including 12 binaries of which only 6 lie in the 2.2--25 arcsec separation range. Two of those six binaries are in our sample, and five are M low-metallicity dwarfs, yielding a binary fraction of 2.33$\pm$2.04\%. This exercise supports our finding that ultracool subdwarfs have a much lower multiplicity fraction than higher mass subdwarfs.

As for theoretical predictions, hydrodynamical simulations predict a multiplicity fraction of 15--25\% and 12\% for subsolar metallicity (0.1\,$Z_{\astrosun}$) M and L dwarfs, respectively \citep{bate14b,bate19a}. The impact of metallicity on the multiplicity appears very limited with fractions of 15--40\% for metal-poor M dwarfs, and  $\sim$10\% for metal-poor L dwarfs, when $Z$\,=\,0.01 $Z_{\astrosun}$. These studies show that our findings are far from these numbers, but those simulations only consider separations below 10\,000\,au, with most of them lower than 1\,000\,au, while our pairs are beyond the upper limits (16\,500--142\,400\,au), except for pair Id 150/150--1 at 1\,360\,au. Therefore, our study is not sensitive to such short separations and we are not able to directly compare or test theoretical predictions.

\begin{table}
 \centering
 \caption[]{Binary fractions obtained in this work.}
 \scalebox{0.9}[0.9]{
 \begin{tabular}{ccccc}
 \hline
\noalign{\smallskip} 
 Sample & [M/H] & Binaries & Sample size & Binary fraction \\
        &   dex     &          &             &       \%      \\
\noalign{\smallskip}        
 \hline
 \hline
\noalign{\smallskip} 
sdM5--L8 & $\leq -0.5$ & 5 & 218 & 2.29$\pm$2.01 \\
\noalign{\smallskip}
sdM5--9.5 & $\leq -0.5$ & 2 & 167 & $0.60_{-0.60}^{+1.17}$ \\ 
\noalign{\smallskip}
sdL & $\leq -0.5$ & 3 & 49 & $6.12_{-6.12}^{+6.93}$\tablefootmark{a} \\
\noalign{\smallskip} 
 \hline
\noalign{\smallskip} 
sdM & --0.5 & 1 & 97 & $1.03_{-1.03}^{+2.02}$ \\ 
\noalign{\smallskip}
esdM & --1 & 1 & 53 & $1.89_{-1.89}^{+3.70}$ \\
\noalign{\smallskip}
usdM & --2 & 0 & 19 & $\leq 5.3$ \\
\noalign{\smallskip} 
 \hline
 \end{tabular}
}
 \label{tab_wide_sdM:binarity}
\tablefoot{\tablefoottext{a}{With doubtful pair Id 190/190--1, the binary fraction can reach 8.16$\pm$8.00\%.}}
\end{table}

\section{Conclusions}
\label{wide_sdM:conclusions}

We present a dedicated search for wide companions to a sample of spectroscopically confirmed M and L subdwarfs. We identified several candidates around six subdwarfs. Based on these findings, we come to the following conclusions:

\begin{enumerate}
\item We did not find low-mass companions to any of the 219 sources of our sample.
\item We did not detect companions colder than L-type sources because of the sensitivity limit of \textit{Gaia} DR2\@. With these data, we are not able to determine the multiplicity fraction of L subdwarfs.
\item We find a metal-poor M-M system, which has been confirmed spectroscopically, composed of \textit{Gaia} DR2 3720832015084722304 (with spectral type sdM5.5 and effective temperature of 3\,000\,K) and \textit{Gaia} DR2 3720832010789680000 (sdM1.5$\pm$0.5 and 3\,600\,K).
\item We find another M-M system, but of solar metallicity, whose spectroscopy leads us to consider it as a bound system. As its metallicity is higher than that of all the other sources in our sample, we did not include it in the analysis of binarity of subdwarfs.
\item We identified four possible M-L systems, and the spectroscopy seems to confirm one of them as bound. This system is composed of ULAS J124104.75-000531.4 and \textit{Gaia} DR2 3695978963488707072 (whose spectral types and effective temperatures are sdL0$\pm$0.5 and 2\,600\,K, and sdM5$\pm$0.5 and 3\,300\,K, respectively), plus one more system confirmed by \cite{zhang19g} composed of \textit{Gaia} DR2 4818823636756117504 and 2MASS J04524567-3608412 (esdM$\pm$0.5 and esdL0$\pm$0.5, and 3\,700\,K and 2\,800\,K respectively). The remaining two systems should be confirmed spectroscopically in the future. This is an interesting result because we extend the known M-L systems from one to two, and probably four. These new systems are important targets to infer the metallicities of the L subdwarfs with higher precision.
\item We infer a frequency of wide systems among sdM5--sdM9.5 of 0.6\small $_{-0.6}^{+1.2}$\% \normalsize for projected physical separations larger than 1\,360\,au (up to 142\,400\,au).
\item We derive a binarity of 1.03\small $_{-1.03}^{+2.02}$\% \normalsize in M subdwarfs (sdM), while the multiplicity of M extreme subdwarfs (esdM) is 1.89\small $_{-1.89}^{+3.70}$\% \normalsize.
\item We did not find any companion to the M ultracool subdwarfs (usdM) in our sample, placing an upper limit on binarity of 5.3\%.
\end{enumerate}
Our study reveals new wide companions around the largest sample of ultracool subdwarfs known to date but is limited in depth to higher mass companions. We plan to look for less massive companions with future multi-epoch
deep surveys like Vera Rubin Large Synoptic Survey telescope \citep{ivezic08a} or in the infrared with upcoming space missions like Euclid \citep{laureijs11,amiaux12,mellier16a} or the Wide Field Infrared Survey Telescope \citep[WFIRST;][]{spergel15}.

\begin{acknowledgements}
This publication makes use of VOSA, developed under the Spanish Virtual Observatory project supported by the Spanish MINECO through grant AYA2017-84089.
VOSA has been partially updated by using funding from the European Union's Horizon 2020 Research and Innovation Programme, under Grant Agreement 776403 (EXOPLANETS-A) .
NL was funded by the Spanish Ministry of Economy and Competitiveness (MINECO) through 
programme number AYA2015-69350-C3-2-P\@.
ZHZ is supported by the Fundamental Research Funds for the Central Universities (Grant No.14380034)
This work is based on observations made with the William Herschel telescope (programme C95;
PI Galvez-Ortiz) operated on the island of La Palma by the Isaac Newton Group of Telescopes 
in the Spanish Observatorio del Roque de los Muchachos of the Instituto de Astrof\'isica de Canarias.
The data presented here were obtained  with ALFOSC, which is provided by the Instituto de Astrof\'isica de Andaluc\'ia (IAA) under a joint agreement with the University of Copenhagen and NOTSA\@.
This research has made use of the \textit{Simbad} and \textit{Vizier} databases, operated at the centre de Donn\'ees Astronomiques de Strasbourg (CDS), and of NASA's Astrophysics Data System Bibliographic Services (ADS). This research has also made use of some of the tools developed as part of the Virtual Observatory.
This work has made use of data from the European Space Agency (ESA) mission $Gaia$ (\url{https://www.cosmos.esa.int/gaia}), processed by the $Gaia$ Data Processing and Analysis Consortium (DPAC, \url{https://www.cosmos.esa.int/web/gaia/dpac/consortium}). Funding for the DPAC has been provided by national institutions, in particular the institutions participating in the $Gaia$ Multilateral Agreement.\\
Funding for the Sloan Digital Sky Survey IV has been provided by the Alfred P. Sloan Foundation, the U.S. Department of Energy Office of Science, and the Participating Institutions. SDSS-IV acknowledges support and resources from the Center for High-Performance Computing at the University of Utah. The SDSS web site is www.sdss.org. SDSS-IV is managed by the Astrophysical Research Consortium for the Participating Institutions of the SDSS Collaboration including the Brazilian Participation Group, the Carnegie Institution for Science, Carnegie Mellon University, the Chilean Participation Group, the French Participation Group, Harvard-Smithsonian Center for Astrophysics, 
Instituto de Astrof\'isica de Canarias, The Johns Hopkins University, Kavli Institute for the Physics and Mathematics of the Universe (IPMU) / University of Tokyo, Lawrence Berkeley National Laboratory, Leibniz Institut f\"ur Astrophysik Potsdam (AIP), Max-Planck-Institut f\"ur Astronomie (MPIA Heidelberg), Max-Planck-Institut f\"ur Astrophysik (MPA Garching), 
Max-Planck-Institut f\"ur Extraterrestrische Physik (MPE), National Astronomical Observatories of China, New Mexico State University, New York University, University of Notre Dame, Observat\'ario Nacional / MCTI, The Ohio State University, Pennsylvania State University, Shanghai Astronomical Observatory, United Kingdom Participation Group, Universidad Nacional Aut\'onoma de M\'exico, University of Arizona, University of Colorado Boulder, University of Oxford, University of Portsmouth, University of Utah, University of Virginia, University of Washington, University of Wisconsin, Vanderbilt University, and Yale University.
This publication makes use of data products from the Two Micron All Sky Survey, which is a joint project of the University of Massachusetts and the Infrared Processing and Analysis 
Center/California Institute of Technology, funded by the National Aeronautics and Space 
Administration and the National Science Foundation.\\
The UKIDSS project is defined in \citep{lawrence07}. UKIDSS uses the UKIRT Wide Field Camera \cite[WFCAM;][]{casali07}. The photometric system is described in \citet{hewett06}, and the calibration is described in \citet{hodgkin09}. The pipeline processing and science archive are described in Irwin et al.\ (2009, in prep) and \citet{hambly08}.\\
This publication makes use of data products from the Wide-field Infrared Survey Explorer, which is a joint project of the University of California, Los Angeles, and the Jet Propulsion Laboratory/California Institute of Technology, and NEOWISE, which is a project of the Jet Propulsion Laboratory/California Institute of Technology. WISE and NEOWISE are funded by the National Aeronautics and Space Administration. \\
\end{acknowledgements}

%
\bibliographystyle{aa.bst}
\bibliography{mnemonic,biblio}

%
\begin{appendix}
\label{wide_sdM:appendix}
\onecolumn

\section{Plots for subdwarfs and wide companions}
\label{wide_sdM:appendix_figures_CMD}

\begin{figure*}[h]
  \caption*{\textbf{Id 11}}
  \vspace{2mm}
  \centering
  \includegraphics[width=0.6\linewidth, angle=0]{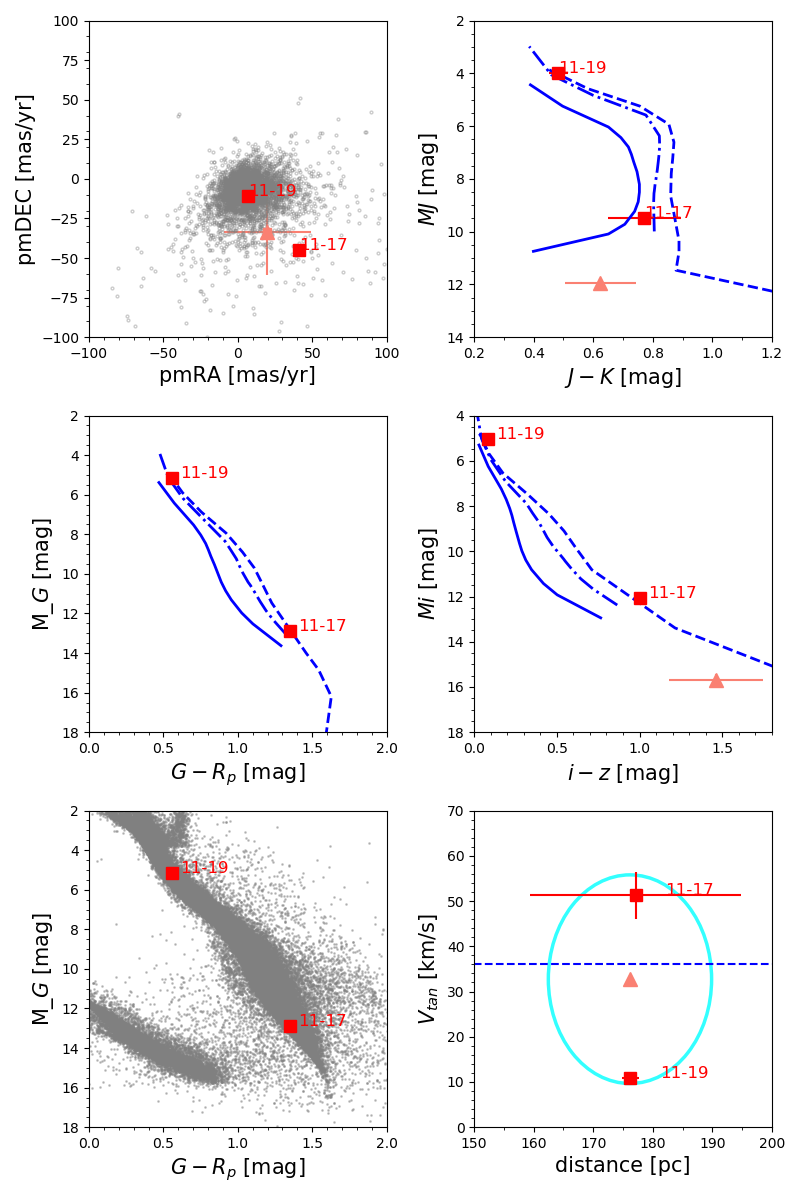}
  \vspace{5mm}
  \caption{PMD (top left panel), CMDs (top right and mid panels), HRD (bottom left panel), and tangential velocity--distance diagram (bottom right) for the target Id 11 and its candidate companions. The filled pink triangle represents the source under study, and the numbered red squares represent the companion candidates. Grey dots in the PMD represent field stars, and in the HRD they are $Gaia$ DR2 sources with parallaxes larger than 10 mas used as a reference. The blue solid, dashed, and dotted lines stand for $[M/H]=-2.0$, $[M/H]=-0.5,$ and $[M/H]=0.0$ \textit{BT-Settl} isochrones in the CMDs. The blue dotted line in the tangential velocity plot marks the value $V_{tan}=36\,km\,s^{-1}$ which is the mean value for field stars \citep{zhang18a}, and the light blue ellipse around Id 11 indicates its values of $V_{tan}\pm \sigma$ and $d \pm \sigma$.}
  \label{fig_wide_sdM:plot_CMD_PM_11}
\end{figure*}

\newpage

\begin{figure*}[h]
  \caption*{\textbf{Id 25}}
  \vspace{10mm}
  \centering
  \includegraphics[width=0.6\linewidth, angle=0]{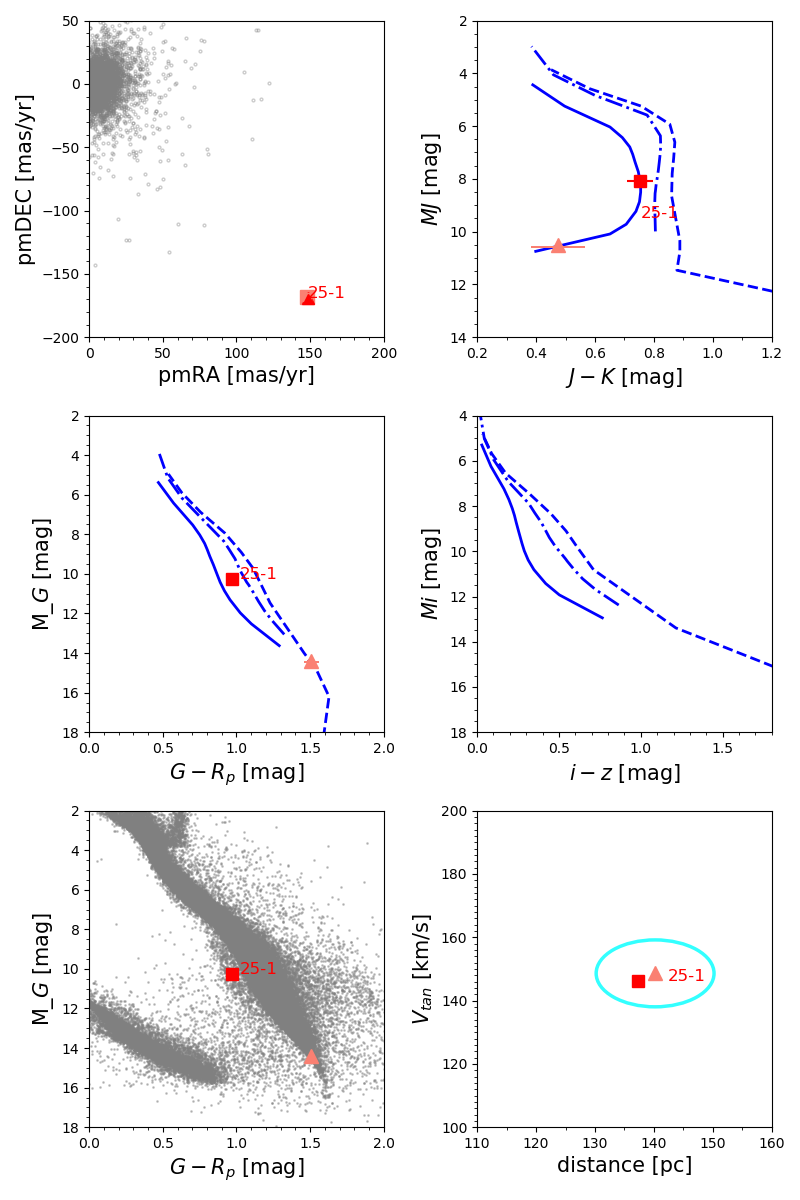}
  \vspace{5mm}
  \caption{PMD (top left panel), CMDs (top right and mid panels), HRD (bottom left panel), and tangential velocity--distance diagram (bottom right) for the target Id 25 and its candidate companion. The filled pink triangle represents the source under study, and the numbered red square represents the companion candidate. Grey dots in the PMD represent field stars, and in the HRD they are $Gaia$ DR2 sources with parallaxes larger than 10 mas used as a reference. The blue solid, dashed and dotted lines stand for $[M/H]=-2.0$, $[M/H]=-0.5$ and $[M/H]=0.0$ \textit{BT-Settl} isochrones in the CMDs. The blue dotted line in the tangential velocity plot marks the value $V_{tan}=36\,km\,s^{-1}$ which is the mean value for field stars \citep{zhang18a} (not visible in the graph), and the light blue ellipse around Id 25 indicates its values of $V_{tan}\pm \sigma$ and $d \pm \sigma$.}
  \label{fig_wide_sdM:plot_CMD_PM_25}
\end{figure*}

\newpage

\begin{figure*}
\centering
\caption*{\textbf{Id 73}}
\vspace{10mm}
  \includegraphics[width=0.6\textwidth]{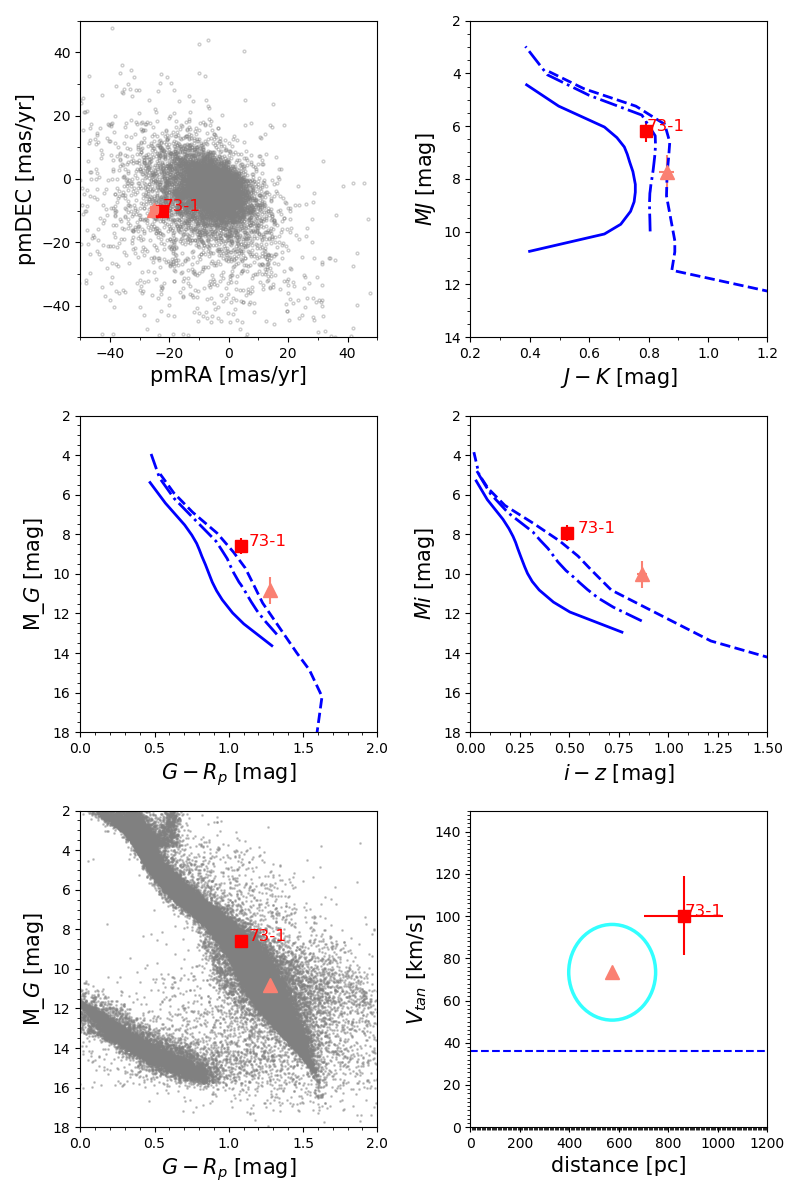}
  \vspace{5mm}
  \caption{PMD (top left panel), CMDs (top right and mid panels), HRD (bottom left panel), and tangential velocity--distance diagram (bottom right) for the target Id 73 and its candidate companion. The filled pink triangle represents the source under study, and the numbered red square represents the companion candidate. Grey dots in the PMD represent field stars, and in the HRD they are $Gaia$ DR2 sources with parallaxes larger than 10 mas used as a reference. The blue solid, dashed and dotted lines stand for $[M/H]=-2.0$, $[M/H]=-0.5$ and $[M/H]=0.0$ \textit{BT-Settl} isochrones in the CMDs. The blue dotted line in the tangential velocity plot marks the value $V_{tan}=36\,km\,s^{-1}$ which is the mean value for field stars \citep{zhang18a}, and the light blue circle around Id 73 indicates its values of $V_{tan}\pm \sigma$ and $d \pm \sigma$.}
   \label{fig_wide_sdM:plot_CMD_PM_73}
\end{figure*}

\newpage

\begin{figure*}[h]
  \caption*{\textbf{Id 89}}
  \vspace{10mm}
  \centering
  \includegraphics[width=0.6\linewidth, angle=0]{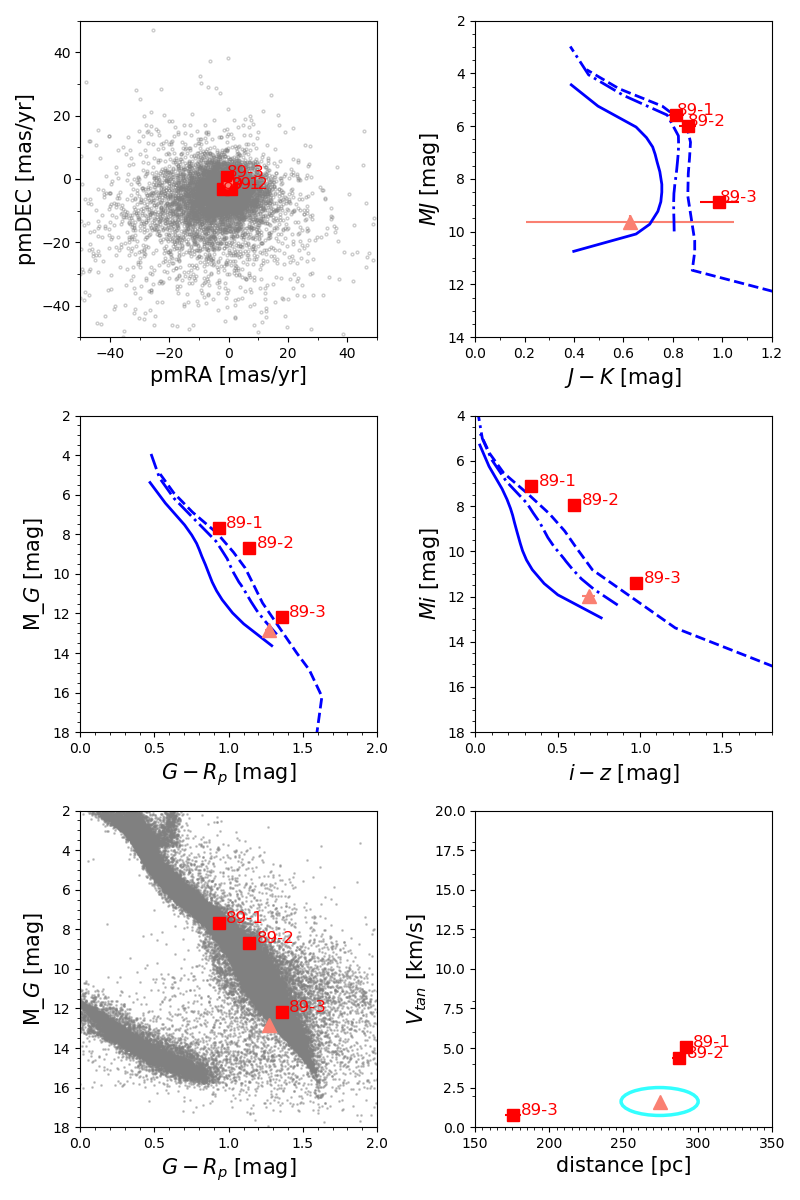}
  \vspace{5mm}
  \caption{PMD (top left panel), CMDs (top right and mid panels), HRD (bottom left panel), and tangential velocity--distance diagram (bottom right) for the target Id 89 and its candidate companions. The filled pink triangle represents the source under study, and the numbered red squares represent the companion candidates. Grey dots in the PMD represent field stars, and in the HRD they are $Gaia$ DR2 sources with parallaxes larger than 10 mas used as a reference. The blue solid, dashed and dotted lines stand for $[M/H]=-2.0$, $[M/H]=-0.5$ and $[M/H]=0.0$ \textit{BT-Settl} isochrones in the CMDs. The blue dotted line in the tangential velocity plot (not visible in the graph) marks the value $V_{tan}=36\,km\,s^{-1}$ which is the mean value for field stars \citep{zhang18a}, and the light blue ellipse around Id 89 indicates its values of $V_{tan}\pm \sigma$ and $d \pm \sigma$.}
   \label{fig_wide_sdM:plot_CMD_PM_89}
\end{figure*}

\newpage

\begin{figure*}[h]
  \caption*{\textbf{Id 107}}
  \vspace{10mm}
  \centering
  \includegraphics[width=0.6\linewidth, angle=0]{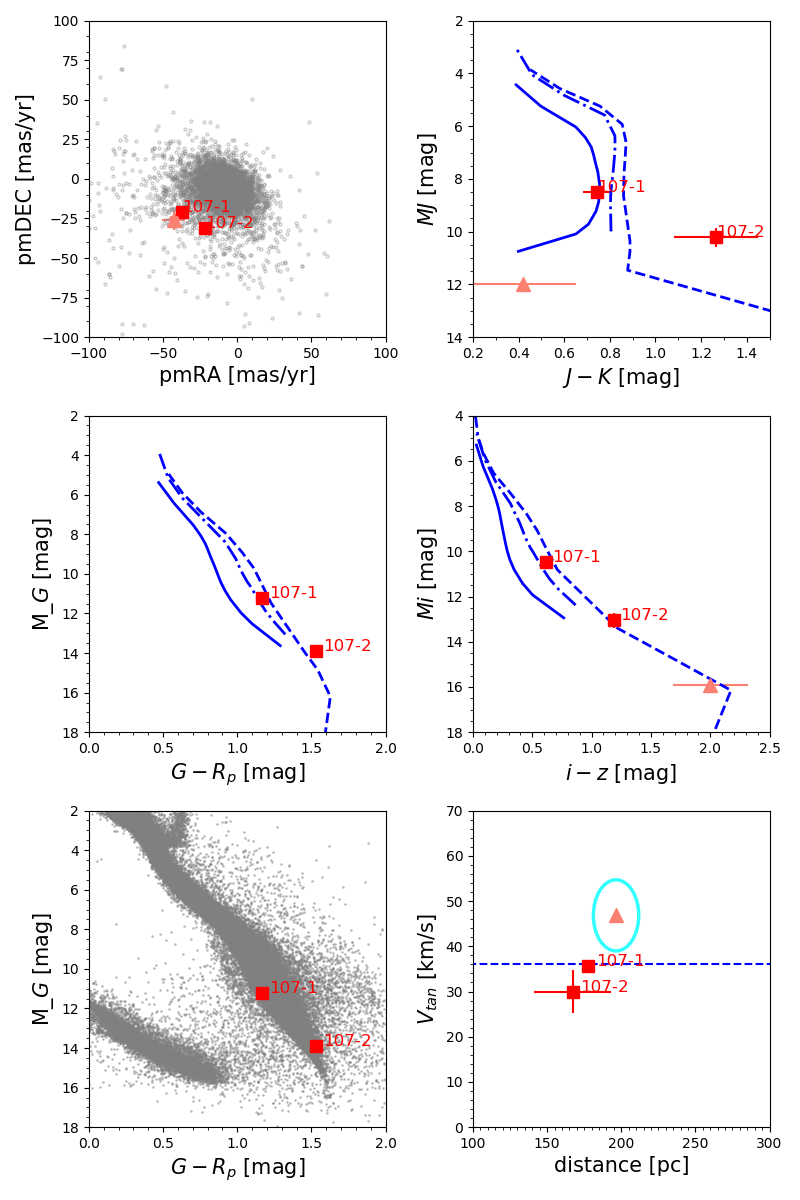}
  \vspace{5mm}
  \caption{PMD (top left panel), CMDs (top right and mid panels), HRD (bottom left panel), and tangential velocity--distance diagram (bottom right) for the target Id 107 and its candidate companions. The filled pink triangle represents the source under study, and the numbered red squares represent the companion candidates. Grey dots in the PMD represent field stars, and in the HRD they are $Gaia$ DR2 sources with parallaxes larger than 10 mas used as a reference. The blue solid, dashed and dotted lines stand for $[M/H]=-2.0$, $[M/H]=-0.5$ and $[M/H]=0.0$ \textit{BT-Settl} isochrones in the CMDs. The blue dotted line in the tangential velocity plot marks the value $V_{tan}=36\,km\,s^{-1}$ which is the mean value for field stars \citep{zhang18a}, and the light blue ellipse around Id 107 indicates its values of $V_{tan}\pm \sigma$ and $d \pm \sigma$.}
  \label{fig_wide_sdM:plot_CMD_PM_107}
\end{figure*}

\newpage

\begin{figure*}[h]
  \caption*{\textbf{Id 126}}
  \vspace{10mm}
  \centering
  \includegraphics[width=0.6\linewidth, angle=0]{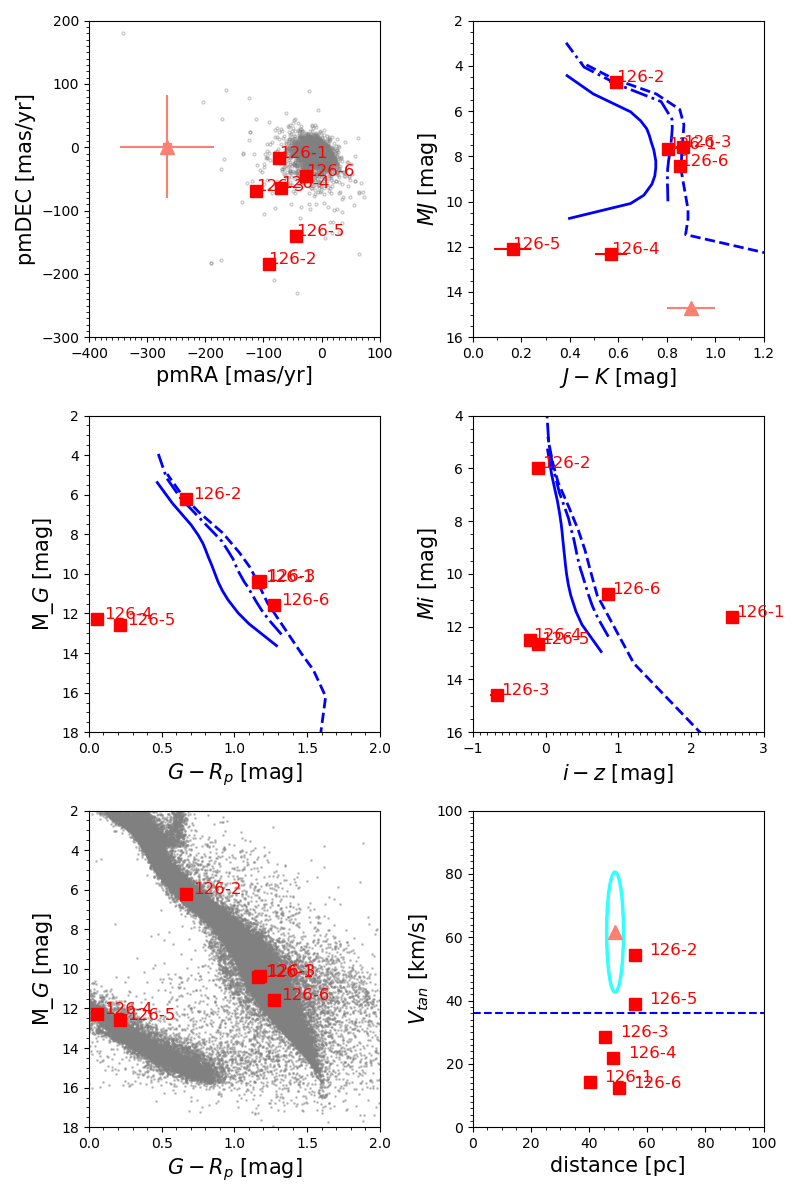}
  \vspace{5mm}
  \caption{PMD (top left panel), CMDs (top right and mid panels), HRD (bottom left panel), and tangential velocity--distance diagram (bottom right) for the target Id 126 and its candidate companions. The filled pink triangle represents the source under study, and the numbered red squares represent the companion candidates. Grey dots in the PMD represent field stars, and in the HRD they are $Gaia$ DR2 sources with parallaxes larger than 10 mas used as a reference. The blue solid, dashed and dotted lines stand for $[M/H]=-2.0$, $[M/H]=-0.5$ and $[M/H]=0.0$ \textit{BT-Settl} isochrones in the CMDs. The blue dotted line in the tangential velocity plot marks the value $V_{tan}=36\,km\,s^{-1}$ which is the mean value for field stars \citep{zhang18a}, and the light blue ellipse around Id 126 indicates its values of $V_{tan}\pm \sigma$ and $d \pm \sigma$.}
  \label{fig_wide_sdM:plot_CMD_PM_126}
\end{figure*}

\newpage

\begin{figure*}[h]
  \caption*{\textbf{Id 128}}
  \vspace{10mm}  
  \centering
  \includegraphics[width=0.6\linewidth, angle=0]{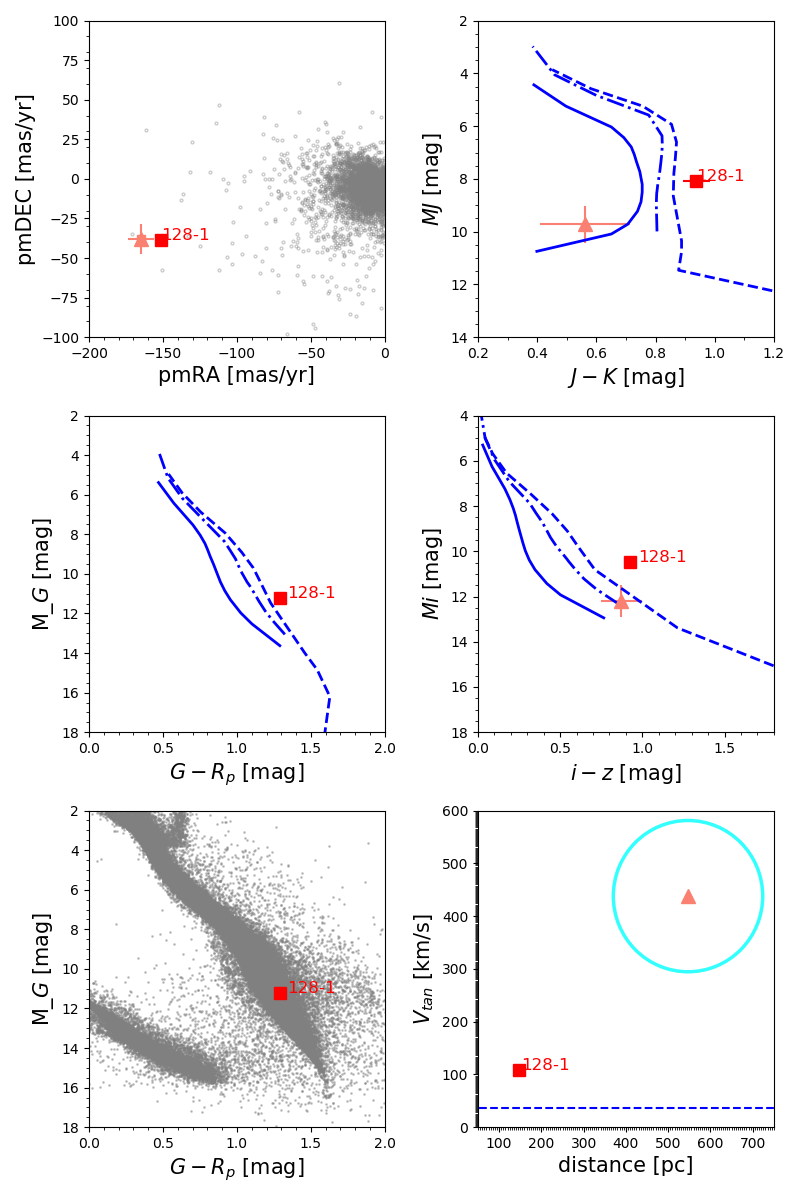}
  \vspace{5mm}  
  \caption{PMD (top left panel), CMDs (top right and mid panels), HRD (bottom left panel), and tangential velocity--distance diagram (bottom right) for the target Id 128 and its candidate companion. The filled pink triangle represents the source under study, and the numbered red square represents the companion candidate. Grey dots in the PMD represent field stars, and in the HRD they are $Gaia$ DR2 sources with parallaxes larger than 10 mas used as a reference. The blue solid, dashed and dotted lines stand for $[M/H]=-2.0$, $[M/H]=-0.5$ and $[M/H]=0.0$ \textit{BT-Settl} isochrones in the CMDs. The blue dotted line in the tangential velocity plot marks the value $V_{tan}=36\,km\,s^{-1}$ which is the mean value for field stars \citep{zhang18a}, and the light blue circle around Id 128 indicates its values of $V_{tan}\pm \sigma$ and $d \pm \sigma$.}
  \label{fig_wide_sdM:plot_CMD_PM_128}
\end{figure*}

\newpage

\begin{figure*}[h]
  \caption*{\textbf{Id 149}}
  \vspace{10mm}
  \centering
  \includegraphics[width=0.6\linewidth, angle=0]{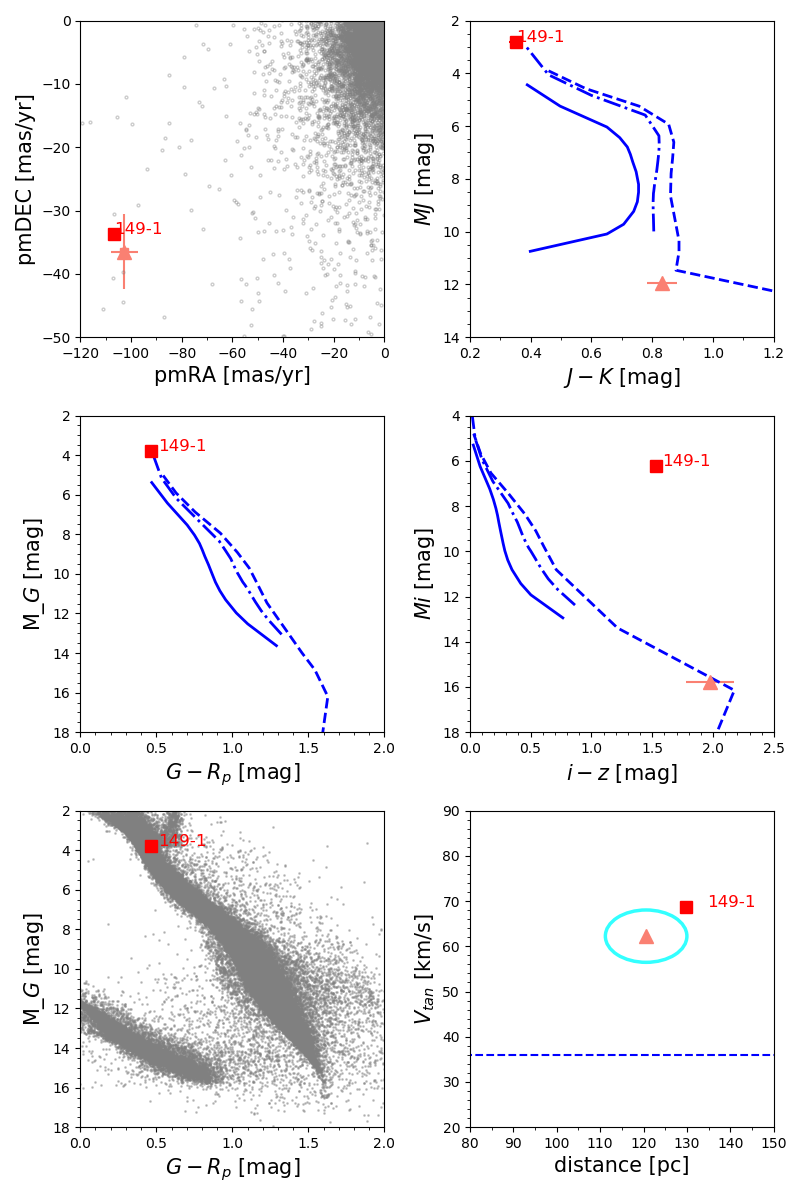}
  \vspace{5mm}
  \caption{PMD (top left panel), CMDs (top right and mid panels), HRD (bottom left panel), and tangential velocity--distance diagram (bottom right) for the target Id 149 and its candidate companion. The filled pink triangle represents the source under study, and the numbered red square represents the companion candidate. Grey dots in the PMD represent field stars, and in the HRD they are $Gaia$ DR2 sources with parallaxes larger than 10 mas used as a reference. The blue solid, dashed and dotted lines stand for $[M/H]=-2.0$, $[M/H]=-0.5$ and $[M/H]=0.0$ \textit{BT-Settl} isochrones in the CMDs. The blue dotted line in the tangential velocity plot marks the value $V_{tan}=36\,km\,s^{-1}$ which is the mean value for field stars \citep{zhang18a}, and the light blue ellipse around Id 149 indicates its values of $V_{tan}\pm \sigma$ and $d \pm \sigma$.}
  \label{fig_wide_sdM:plot_CMD_PM_149}
\end{figure*}

\newpage

\begin{figure*}[h]
  \caption*{\textbf{Id 190}}
  \vspace{10mm}
  \centering
  \includegraphics[width=0.6\linewidth, angle=0]{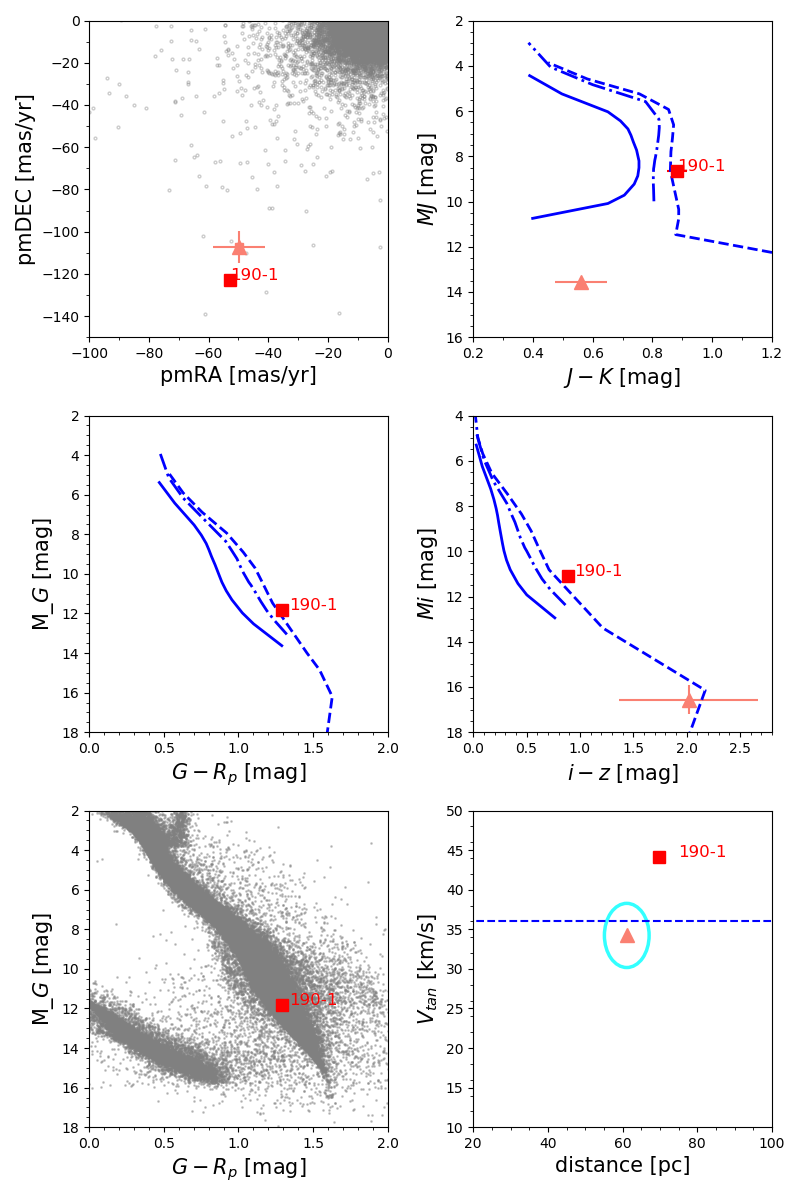}
  \vspace{5mm}
  \caption{PMD (top left panel), CMDs (top right and mid panels), HRD (bottom left panel), and tangential velocity--distance diagram (bottom right) for the target Id 190 and its candidate companion. The filled pink triangle represents the source under study, and the numbered red square represents the companion candidate. Grey dots in the PMD represent field stars, and in the HRD they are $Gaia$ DR2 sources with parallaxes larger than 10 mas used as a reference. The blue solid, dashed and dotted lines stand for $[M/H]=-2.0$, $[M/H]=-0.5$ and $[M/H]=0.0$ \textit{BT-Settl} isochrones in the CMDs. The blue dotted line in the tangential velocity plot marks the value $V_{tan}=36\,km\,s^{-1}$ which is the mean value for field stars \citep{zhang18a}, and the light blue ellipse around Id 190 indicates its values of $V_{tan}\pm \sigma$ and $d \pm \sigma$.}
  \label{fig_wide_sdM:plot_CMD_PM_190}
\end{figure*}

\newpage

\begin{figure*}[h]
  \caption*{\textbf{Id 213}}
  \vspace{10mm}
  \centering
  \includegraphics[width=0.6\linewidth, angle=0]{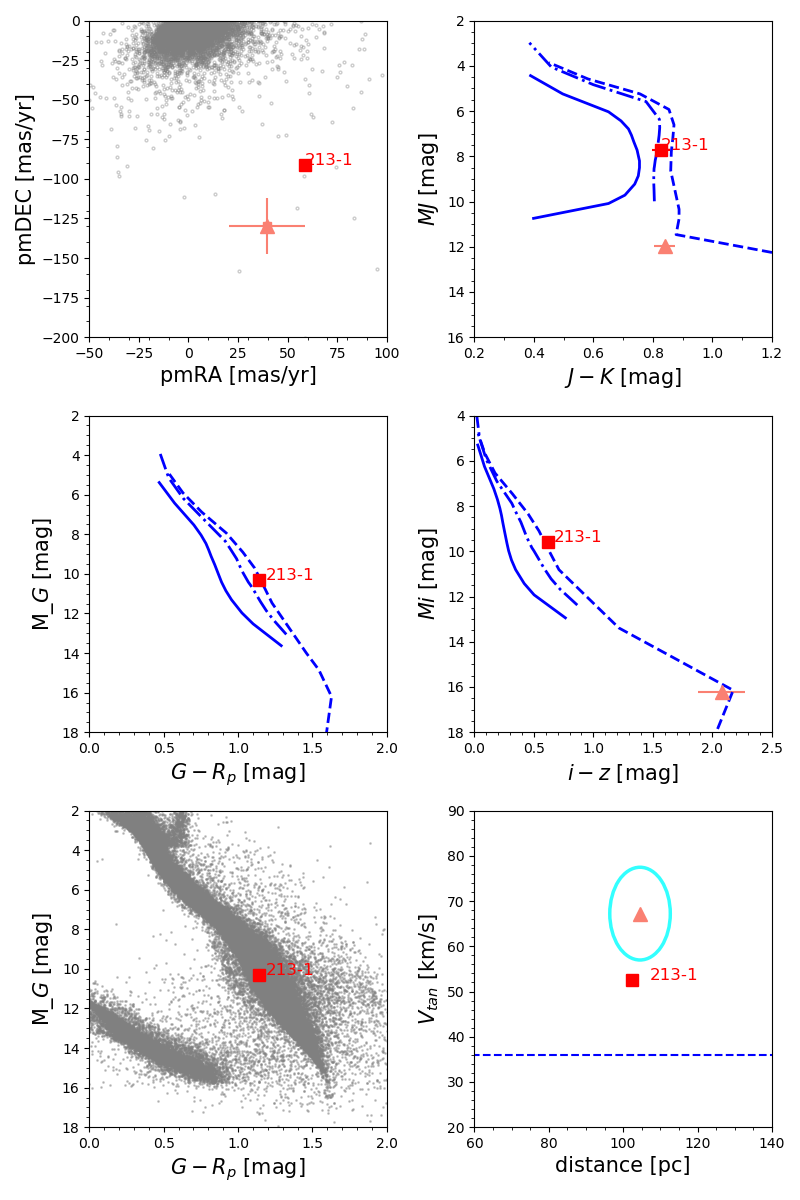}
  \vspace{5mm}
  \caption{PMD (top left panel), CMDs (top right and mid panels), HRD (bottom left panel), and tangential velocity--distance diagram (bottom right) for the target Id 213 and its candidate companion. The filled pink triangle represents the source under study, and the numbered red square represents the companion candidate. Grey dots in the PMD represent field stars, and in the HRD they are $Gaia$ DR2 sources with parallaxes larger than 10 mas used as a reference. The blue solid, dashed and dotted lines stand for $[M/H]=-2.0$, $[M/H]=-0.5$ and $[M/H]=0.0$ \textit{BT-Settl} isochrones in the CMDs. The blue dotted line in the tangential velocity plot marks the value $V_{tan}=36\,km\,s^{-1}$ which is the mean value for field stars \citep{zhang18a}, and the light blue ellipse around Id 213 indicates its values of $V_{tan}\pm \sigma$ and $d \pm \sigma$.}
  \label{fig_wide_sdM:plot_CMD_PM_213}
\end{figure*}

\newpage

\centering
\begin{figure*}[h]
  \caption*{\textbf{Id 215}}
  \vspace{10mm}
  \centering
  \includegraphics[width=0.6\linewidth, angle=0]{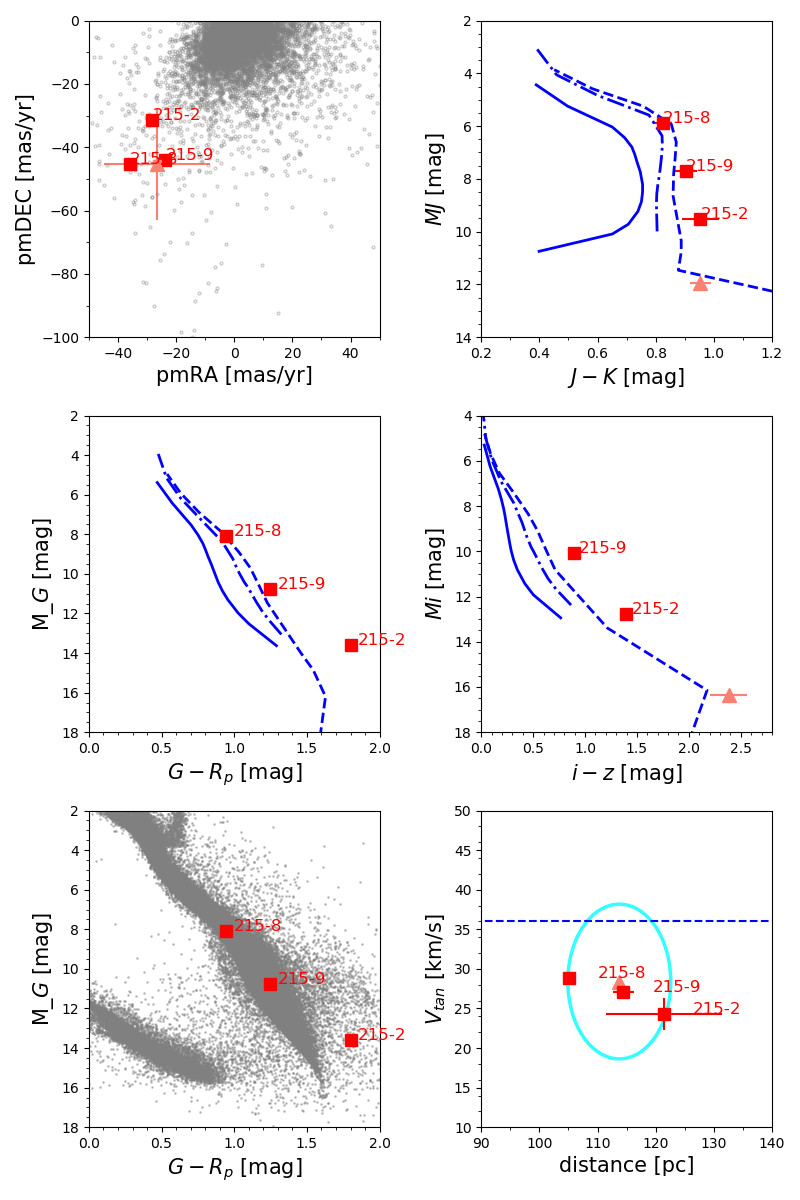}
  \vspace{5mm}
  \caption{PMD (top left panel), CMDs (top right and mid panels), HRD (bottom left panel), and tangential velocity--distance diagram (bottom right) for the target Id 215 and its candidate companions. The filled pink triangle represents the source under study, and the numbered red squares represent the companion candidates. Grey dots in the PMD represent field stars, and in the HRD they are $Gaia$ DR2 sources with parallaxes larger than 10 mas used as a reference. The blue solid, dashed and dotted lines stand for $[M/H]=-2.0$, $[M/H]=-0.5$ and $[M/H]=0.0$ \textit{BT-Settl} isochrones in the CMDs. The blue dotted line in the tangential velocity plot marks the value $V_{tan}=36\,km\,s^{-1}$ which is the mean value for field stars \citep{zhang18a}, and the light blue ellipse around Id 215 indicates its values of $V_{tan}\pm \sigma$ and $d \pm \sigma$.}
  \label{fig_wide_sdM:plot_CMD_PM_215}
\end{figure*}

\newpage
\clearpage
\section{Additional table}
\label{wide_sdM:additional_table}

\begin{table*}[h]
 \centering
 \caption{Summary of the companion candidates assessment.}
 \scalebox{0.85}[0.85]{
 \begin{tabular}{lcccccccc}
\hline
\noalign{\smallskip}
 Id & PMD & CMD & CMD & CMD & HRD & Kinematics & $V_{Tan}$ & Candidate \\
  & & $M_j/J-K$ & $M_G/G-RP$ & $M_i/i-z$ & & & & \\ 
 \noalign{\smallskip}  
\hline
\hline
\noalign{\smallskip}
 11-1 & No & - & - & - & - & No & - & No \\
 11-2 & No & - & - & - & - & - & - & No \\ 
 11-3 & No & - & - & - & - & - & - & No \\ 
 11-4 & No & - & - & - & - & - & - & No \\ 
 11-5 & No & - & - & - & - & - & - & No \\ 
 11-6 & No & - & - & - & - & - & - & No \\
 11-7 & No & - & - & - & - & No & - & No \\ 
 11-8 & No & - & - & - & - & No & - & No \\ 
 11-9 & No & - & - & - & - & No & - & No \\ 
 11-10 & No & - & - & - & - & No & - & No \\
 11-11 & No & - & - & - & - & - & - & No \\ 
 11-12 & No & - & - & - & - & - & - & No \\ 
 11-13 & No & - & - & - & - & - & - & No \\ 
 11-14 & No & - & - & - & - & No & - & No \\ 
 11-15 & No & - & - & - & - & No & - & No \\ 
 11-16 & No & - & - & - & - & - & - & No \\ 
 11-17 & Yes & Yes & ? & Yes & Yes & - & Yes & Yes \\
 11-18 & No & - & - & - & - & - & - & No \\ 
 11-19 & Yes & Yes & ? & Yes & No & No & No & No \\
 11-20 & No & - & - & - & - & No & - & No \\ 
 11-21 & No & - & - & - & - & - & - & No \\ 
 11-22 & No & - & - & - & - & - & - & No \\ 
 11-23 & No & - & - & - & - & No & - & No \\ 
 11-24 & No & - & - & - & - & - & - & No \\ 
 11-25 & No & - & - & - & - & - & - & No \\ 
 11-26 & No & - & - & - & - & No & - & No \\ 
 11-27 & No & - & - & - & - & - & - & No \\ 
 11-28 & No & - & - & - & - & - & - & No \\ 
 11-29 & No & - & - & - & - & - & - & No \\ 
 11-30 & No & - & - & - & - & - & - & No \\ 
 11-31 & No & - & - & - & - & - & - & No \\ 
 11-32 & No & - & - & - & - & - & - & No \\ 
 25-1 & Ýes & Yes & Yes & - & Yes & - & Yes & Yes \\
 73-1 & Yes & Yes & Yes & Yes & Yes & - & Yes & Yes \\
 89-1 & ? & Yes & No & No & No & - & No & No \\
 89-2 & ? & Yes & No & No & No & - & No & No \\
 89-3 & ? & Yes & No & No & No & - & No & No \\
 107-1 & Yes & Yes & ? & Yes & Yes & - & Yes & Yes \\
 107-2 & ? & No & ? & Yes & No & - & No & No \\ 
 126-1 & No & ? & Yes & No & Yes & No & No & No \\ 
 126-2 & No & ? & Yes & Yes & Yes & No & Yes & No \\ 
 126-3 & No & ? & Yes & No & Yes & - & No & No \\ 
 126-4 & No & ? & No & No & No & - & No & No \\ 
 126-5 & No & No & No & No & Yes & No & No & No \\ 
 126-6 & No & ? & Yes & Yes & Yes & - & No & No \\ 
 128-1 & ? & No & ? & No & Yes & - & No & No \\
 149-1 & ? & No & ? & No & Yes & No & No & No \\
 150-1 & Yes & Yes & Yes & Yes & Yes & - & Yes & Yes \\
 190-1 & Yes & No & ? & Yes & Yes & - & Yes & Yes ? \\
 213-1 & No & ? & ? & Yes & Yes & No & No & No \\
 215-1 & No & Yes & ? & Yes & Yes & - & No & No \\ 
 215-2 & Yes & Yes & ? & ? & No & - & Yes & No \\ 
 215-3 & No & Yes & ? & No & Yes & - & No & No \\ 
 215-4 & No & No & ? & No & No & - & No & No \\ 
 215-5 & No & Yes & ? & No & Yes & - & No & No \\ 
 215-6 & No & ? & ? & No & No & - & No & No \\ 
 215-7 & No & Yes & ? & No & Yes & - & No & No \\ 
 215-8 & Yes & Yes & ? & No & Yes & No & Yes & No \\
 215-9 & Yes & Yes & ? & No & Yes & - & Yes & No \\ 
 215-10 & No & No & ? & No & No & - & No & No \\ 
 215-11 & No & No & ? & No & No & No & No & No \\ 
 215-12 & No & Yes & ? & Yes & Yes & - & No & No \\ 
\noalign{\smallskip}
\hline
\end{tabular}
 }
 \label{tab_wide_sdM:proposed_candidates}
\end{table*}

\end{appendix}
\end{document}